\documentclass[aps,prb,amsmath,amssymb,twocolumn]{revtex4-2}

\usepackage{graphicx}
\usepackage{adjustbox}
\usepackage{bm}
\usepackage{color}
\usepackage{braket}
\usepackage{standalone}
\usepackage{multirow}
\usepackage{tikz}
\usepackage{mathrsfs}
\usepackage{dsfont}
\usepackage[colorlinks,bookmarks=true,citecolor=blue,linkcolor=red,urlcolor=blue]{hyperref}

\newcommand{\bea}{\begin{eqnarray}}
\newcommand{\eea}{\end{eqnarray}}
\definecolor{dgreen}{rgb}{0.1,0.5,0.1}
\definecolor{red}{rgb}{1,0,0}

\usepackage{bbm}
\usepackage[caption=false]{subfig}
\usepackage{floatrow}
\usepackage{stackrel}
\usepackage{wrapfig}

\usepackage{tikz-feynman}
\usepackage{tikz}
\usetikzlibrary{positioning}

\begin{document}

\title{Mesoscopic transport signatures of disorder-induced non-Hermitian phases}
\author{Benjamin Michen}
\email{benjamin.michen@tu-dresden.de}
\author{Jan Carl Budich}
\email{jan.budich@tu-dresden.de}
\affiliation{Institute of Theoretical Physics${\rm ,}$ Technische Universit\"{a}t Dresden and W\"{u}rzburg-Dresden Cluster of Excellence ct.qmat${\rm ,}$ 01062 Dresden${\rm ,}$ Germany}
\date{\today}

\begin{abstract}
We investigate the impact on basic quantum transport properties of disorder-induced exceptional points (EPs) that emerge in a disorder-averaged Green's function description of two-dimensional (2D) Dirac semimetals with spin- or orbital-dependent potential scattering. Remarkably, we find that EPs may promote the nearly vanishing conductance of a finite sample at the Dirac point to a sizable value that increases with disorder strength. This striking behavior exhibits a strong directional anisotropy that is closely related to the Fermi arcs connecting the EPs. We corroborate our results by numerically exact simulations, thus revealing the fingerprints of characteristic non-Hermitian spectral features also on the localization properties of the considered systems. Finally, several candidates for the experimental verification of our theoretical analysis are discussed, including disordered electronic square-net materials and cold atoms in spin-dependent optical lattices.
\end{abstract}

	\maketitle

 \section{Introduction}
 The study of disorder in quantum materials has led to the discovery of numerous intriguing phenomena, prominently including metal-insulator transitions and disorder induced topological states \cite{Anderson_Loc_1, Anderson_Loc_2, MB_localization_1, Disorder_Transition_1, Disorder_Transition_2, Disorder_Transition_Topological}. Quite generally, scattering off random impurity potentials tends to inhibit bulk transport by localizing the current-carrying Bloch states of a clean system \cite{Anderson_Loc_1, Anderson_Loc_2}. In this context, the recently proposed framework of many-body localization has aimed at generalizing the basic picture of Anderson localization beyond the independent electron approximation \cite{MB_localization_1, MB_localization_2, MB_localization_3}.
 
Despite the Hermitian character of the microscopic Hamiltonian, non-Hermitian (NH) physics naturally emerges in the description of disordered systems due to the presence of a self-energy $\Sigma = \Sigma_H + i \Sigma_A$ that accounts for impurity scattering in the disorder averaged Green's function. As a simplest scenario, the anti-Hermitian part $\Sigma_A$ of the self-energy just gives a scattering-induced finite lifetime to the stationary Bloch states of the free system, while the Hermitian part $\Sigma_H$ amounts to a correction of the free band structure described by the Bloch Hamiltonian $H_0(\mathbf k)$ with the lattice momentum $\mathbf k$. However, as soon as $\Sigma_A$ does not commute with $H_0(\mathbf k)$, remarkable phenomena unique to NH matrices may occur, such as the formation of topologically stable exceptional points (EPs) \cite{EPDisorder1, EPDisorder2, EPDisorder3} representing the NH counterpart of topological semimetals known from the Hermitian realm, and thus an example of NH topological phases \cite{Rudner2009, Zeuner2015, lee, Zhou2018, Lieu2018, Longhi2018, BBC, yaowang, Imhof2018, gong, Kawabata2019, Budich2019, Yoshida2019, EPringExp, KawabataEP2019, KunstDwivedi2019, Song2019, QdynBBC, SzameitScience2020, Budich2020, EP_Purcell_Enhancement, Sensing_Microcavity, review, Feinberg1997_Ref_B, Mondragon2013_Ref_B, Ostahie2021_Ref_B, Brandenbourger2019,Ghatak2020, Kozii2017,Bergholtz2019,EPManyBody,McClarty2019,Budich2020,Rausch2021} (see Ref.~\cite{review} for an overview).

 \floatsetup[figure]{style=plain,subcapbesideposition=top} 
\begin{figure}[htp!]	 
\sidesubfloat[]{\includegraphics[trim={0cm 16cm 3cm 0cm}, width=0.95\linewidth]{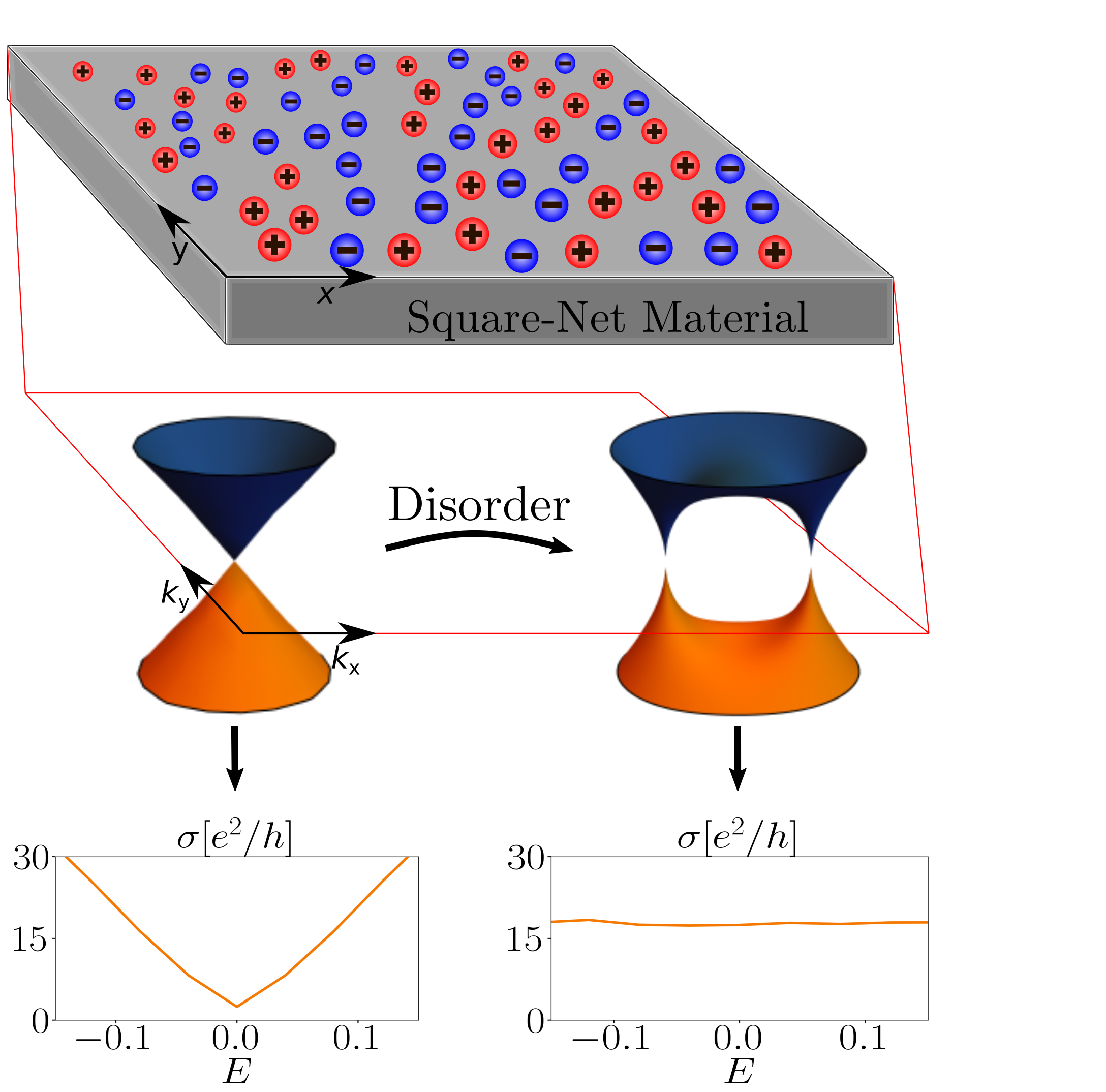}}\\
\sidesubfloat[]{\includegraphics[trim={-1.7cm -2cm 1.7cm 0cm}, clip, width=0.95\linewidth]{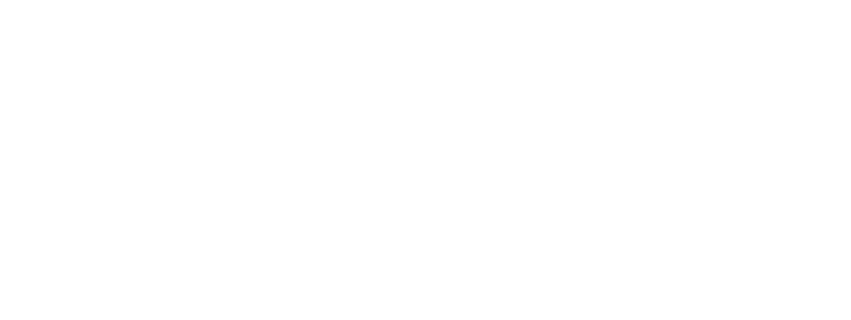} \label{fig1b} }\\
\sidesubfloat[]{\includegraphics[trim={-1.7cm 1cm 1.7cm 0cm}, clip, width=0.95\linewidth]{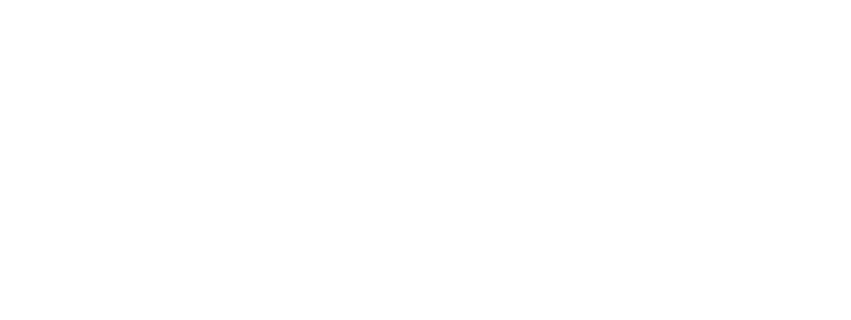} \label{fig1c} }
\caption{(a) Illustration of a disordered two-dimensional (2D) material, which hosts  a semimetallic Dirac dispersion in the clean limit, e.g. realized as a square-net material \cite{SQN_Dirac_1, SQN_Dirac_2} with positive and negative ions adsorbed to the surface. 
(b) For orbital-selective disorder, the Dirac cones may split into topologically stable exceptional points exhibiting a characteristic square-root-like dispersion (absolute value of complex energy is shown).
(c) For a sample of mesoscopic size, the vanishing zero-energy transmission seen in the clean regime (left panel) may be promoted to a finite value (right panel) that increases with disorder strength. \label{illustration_setup}}
\end{figure}

In this work, we study the occurrence of disorder-induced exceptional phases in two-dimensional Dirac semimetals with spin- or orbital-dependent impurity scattering (see Fig.~\ref{illustration_setup}). To this end, we compute the effective NH Hamiltonian
\begin{align}
H_{e}(\bm k) = H_0(\bm k) + \Sigma(\bm k,\omega=0),
\label{eqn:heff}
\end{align}
from exact numerical data on the disorder averaged Green's function \cite{MBQT} at the Fermi Energy $(\omega = 0)$ (cf. Section~\ref{Sec:Spec_Prop} below). 

While previous work on disorder-induced NH phases has mostly concentrated on spectral properties, our main focus is on the mesoscopic quantum transport properties of exceptional NH phases in a simple two-terminal setting \cite{Datta}. Interestingly, we find that the fast exponential decay of the zero-energy conductance with system size, which is caused by the vanishing densitiy of states (DOS) at the nodal points in the clean regime, can be prolonged extensively in the presence of disorder-induced EPs. As a result, the nearly vanishing conductance of a clean mesoscopic sample with the Fermi energy at the Dirac point may be promoted to a sizable value that increases with disorder strength (see Fig.~\ref{illustration_setup}(c), and Fig.~\ref{Transmission_length_arc} below, respectively). Both the directional anisotropy and the value of the transmission are found to be closely related to the orientation and length of the Fermi arcs connecting the EPs as well as the associated quasiparticles with arbitrarily prolonged lifetime that emerge in a certain parameter range. Comparing our results to a conventional disordered phase without EPs, where transport is overall strongly damped and the exponential decay of the zero-energy conductance occurs even faster than in the clean regime, our findings presented in Section~\ref{Sec:Transport} provide clear fingerprints of disorder-induced EPs on basic transport properties.

We find that the aforementioned quasiparticles with largely extended lifetime in the exceptional phase are tied to exact Bloch wave eigenstates that emerge in the limit of orbital-selective disorder and persist for any perturbation amplitude. In section~\ref{Sec:Surviving_Bloch_Modes}, this scenario is generalized to a theorem that guarantees the presence of Bloch eigenstates, whenever only a sufficiently constrained subset of the orbitals are affected by disorder in an arbitrary tight binding model.

Additionally, we investigate the localization properties of the different phases in section~\ref{Sec:Localization}, and find hallmarks of anomalous localization (i.e. slower than exponential decay of the eigenfunctions)  in the exceptional phase for system sizes that are amenable to numerical studies. 

Finally, in Section~\ref{Sec:Platforms}, we outline several candidate platforms for the experimental verification of our theoretical analysis such as magnetically disordered 2D electron systems, square net materials, which refer to a class of three-dimensional (3D) multiatomic crystals with square-shaped 2D sublattices \cite{SQN_Dirac_1, SQN_Dirac_2} hosting a Dirac-like dispersion, and cold atoms in spin-dependent optical lattices.

 \section{Model and methods} 
\begin{figure}[htp!]
  \centering
  \includegraphics[trim={0cm 0cm 0cm 0cm}, clip, width=0.95\linewidth]{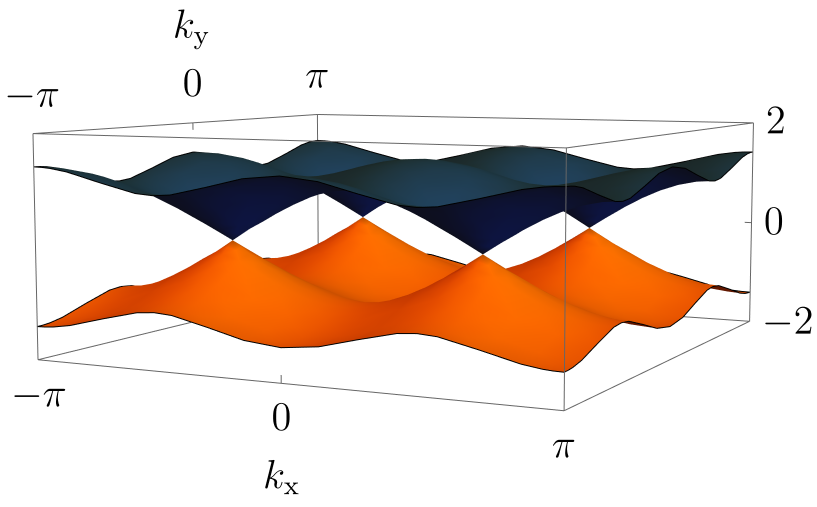}
\caption{Band structure of the free Hamiltonian $H_0$ (see Eq.~(\ref{H_0})) exhibiting 4 Dirac cones with the parameters set to $t_\mathrm{x} = 0.5$,
$t_\mathrm{z} = 0.5$. \label{spec_H0} }
\end{figure}

We consider a conceptually simple square lattice model of a  Dirac semimetal in two spatial dimensions (2D), specified by the Hamiltonian $H_0$ in second-quantized tight-binding form as

\begin{align}
H_0 =&\sum_{\bm j}  \left[ t_\mathrm{x}  \psi_{\bm j+ \bm \delta_\mathrm{y}}^\dagger 
\sigma_\mathrm{x}  \psi_{\bm j} 
+ t_\mathrm{z}   \psi_{\bm j+ \bm \delta_\mathrm{x}}^\dagger \sigma_\mathrm{z}
 \psi_{\bm j}   \right]  + \mathrm{h.c.}  \label{H_0}
\end{align}
with $\bm j = (j_\mathrm{x}, j_\mathrm{y})$ as well as $\bm \delta_\mathrm{x} = (1,0)$, $ \bm \delta_\mathrm{y} = (0,1)$, and $\sigma_\mu,~\mu=\mathrm{x,y,z}$ denoting the standard Pauli matrices. Length is measured in units of the lattice constant,
 energy in terms of the spin-dependent nearest neighbor hopping chosen as $t_\mathrm{x} = t_\mathrm{z} = 0.5$, and the
spinors of field-operators $ \psi_{\bm j}^\dagger = (\psi_{\bm j, \uparrow}^\dagger, \psi_{\bm j, \downarrow}^\dagger) $ create a fermion in unit cell $\bm j$ with spin $\uparrow$ ($\downarrow$). In reciprocal space, the model is characterized by the  Bloch Hamiltonian $H_0(\bm k) = {\bm d_R(\bm  k) \cdot  \bm \sigma}$ with ${\bm d}_R(\bm  k) =  (\mathrm{cos}(k_\mathrm{y}), 0,\mathrm{cos}(k_\mathrm{x
}))$, the spectrum of which exhibits four nodal points in the first Brillouin zone (see Fig.~\ref{spec_H0}). 

Random disorder with a spin-dependent structure is introduced to the system by means of the operator $V$, given in real space representation as
\begin{align}
V = &\sum_{\bm j} a_{\bm j} \left[ \psi_{\bm j}^\dagger \left( 
s_0 \sigma_\mathrm{0} + \gamma \left(\sin(\phi) \sigma_\mathrm{x} + \cos(\phi)  \sigma_\mathrm{z} \right) \right) \psi_{\bm j} 
 + \mathrm{h.c.} \right ],  \label{disorder}
\end{align}
with $s_0, \gamma, \phi \in \mathbb R$. The uncorrelated random amplitudes $\{ a_{\bm j} \}$ are drawn from the box distribution on the real interval $[- \alpha, \alpha]$, so the overall disorder strength scales with $\alpha$. Depending on the choice of disorder parameters, an exceptional or an ordinary phase may emerge.  In Section~\ref{Sec:Platforms}, we outline several possible platforms for the experimental implementation of our model. We again stress that the total Hamiltonian $H= H_0 + V$ is Hermitian, and that non-Hermitian physics only enters via the self-energy in the Green's function description discussed in the following. 

To describe the electronic excitations close to the Fermi level in disordered systems, we employ an effective NH Hamiltonian $H_e(\bm k) = H_0(\bm k) + \Sigma(\bm k,\omega=0)$  (cf.~Eq.~\ref{eqn:heff}), which we obtain from numerically exact data on the disorder-averaged Green's function. Starting from the propagator $G^\mathrm{R}_{\bm k,\bm k'}(t-t') =-i\Theta(t-t') \langle 0|\{ c_{\bm k}(t), c^\dagger_{\bm k'}(t')\}|0 \rangle$, where  
$c_{\bm k}^\dag(t) = (c_{\bm k,\uparrow}^\dag(t), c_{\bm k,\downarrow}^\dag(t))$ is the Heisenberg picture spinor of creation operators in reciprocal space, a Fourier transform with respect to $(t-t')$ yields the retarded Green's function in frequency space. Averaging over the random disorder terms restores translational invariance and renders the the resulting disorder-averaged Green's function $G^{\mathrm{R,av}}_{\bm k}$ block diagonal in momentum space (see Appendix~\ref{perturbation_theory} for details). The blocks $G^{\mathrm{R,av}}_{\bm k}(\omega) = [\mathbbm{1}(\omega + i \eta) - H_0(\bm k) - \Sigma(\bm k,\omega)]^{-1}$, where $\eta >0$ is an infinitesimal regularization, determine the self-energy correction $\Sigma(\bm k,\omega)$ that enters the effective Hamiltonian $H_e(k)$. In the following, we calculate $G^{\mathrm{R,av}}_{\bm k}$ in a numerically exact fashion for a system size of $100\times100$ sites.

\section{Spectral properties} \label{Sec:Spec_Prop}
In this section, we briefly discuss the phenomenon of EPs and show how a conventional disordered phase without EPs or an exceptional phase featuring EPs  emerges in our model system specified by Eqs.~(\ref{H_0},~\ref{disorder}).

\subsection{Exceptional points}

For parameter-dependent NH matrices, EPs refer to points in parameter space at which not only the eigenvalues but also the corresponding eigenvectors coalesce, thereby rendering the matrix non-diagonalizable \cite{Kato, BerryDeg, Heiss}. Since all Hermitian matrices  possess a complete set of eigenvectors, this is a genuinely NH phenomenon. For an overview on the role of EPs in contemporary physics, see, e.g.,  Refs.~\cite{review, EPreview}. 

 \floatsetup[figure]{style=plain,subcapbesideposition=top} 
\begin{figure*}[htp!]	 
  \centering 
\sidesubfloat[]{\includegraphics[trim={0cm 0cm 0cm 0.3cm}, width=\linewidth]{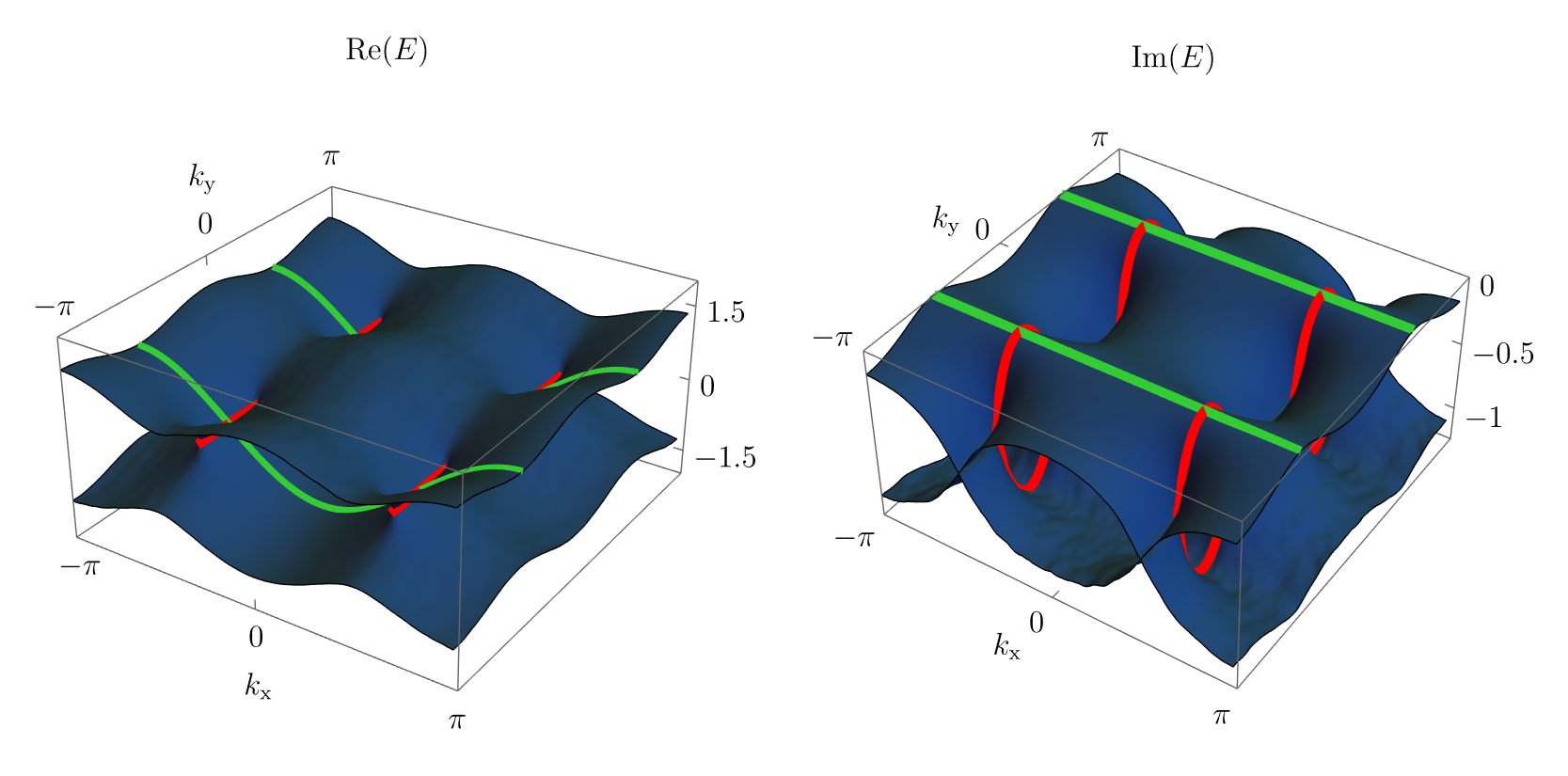} \label{spec_Heff}}\\
\sidesubfloat[ ]{\includegraphics[trim={0cm 0cm 3cm 1cm}, width=1.05\linewidth]{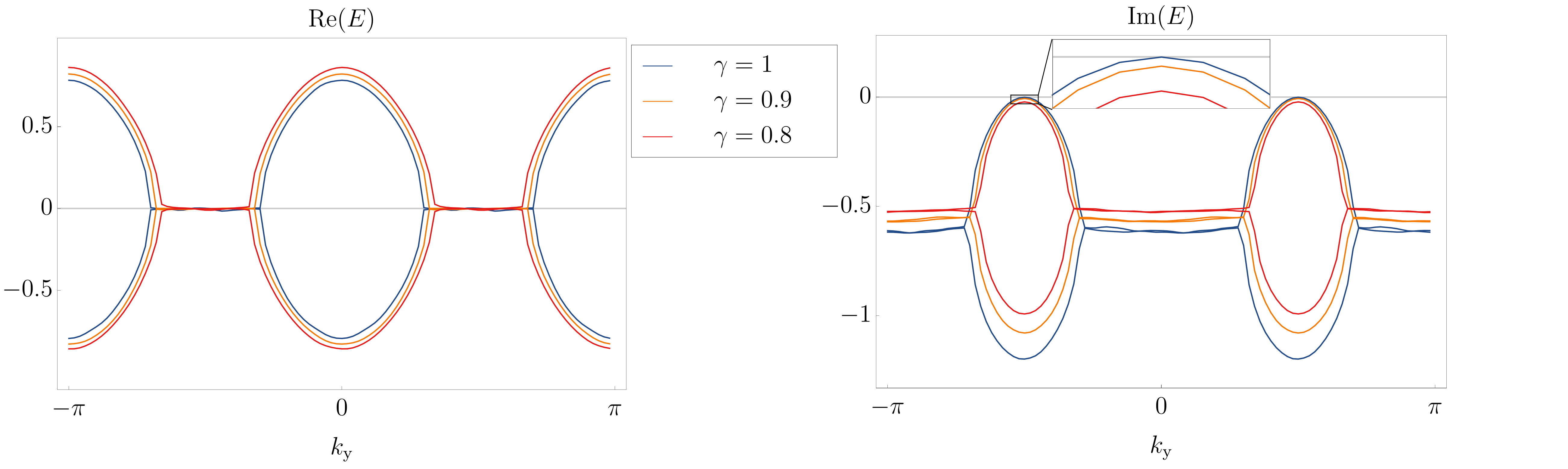}
\label{spec_Fermi_arc_cut}} 
\caption{(a) Numerically exact spectrum of $H_e(k)$ in the exceptional phase. Nodal points of the clean are split into EPs connected by Fermi arcs (marked by red lines). Lines of vanishing imaginary part (marked in green) entail propagating quasiparticles with infinite lifetime. The group velocity along the green lines is parallel to the  $k_\mathrm{x}$-axis, which leads to a directional dependence of transport signatures. Disorder parameters are $\alpha =1.5$, $s_0 = 1.0$, $\gamma= 1$, $\phi = 0$
(b) Cut along the line where  $k_\mathrm{x} = \frac{ \pi}{2}$, showing the profile of the two  Fermi arcs in the $(+,+)$ and $(+,-)$ quadrants of Fig.~\ref{spec_Heff}. Note that the imaginary "Fermi bubbles" touch the real axis. Disorder parameters are $\alpha =1.5$, $s_0 = 1.0$, $\phi = 0$. Different colors correspond to $\gamma=1$ (blue), $\gamma=0.9$ (orange), and $\gamma=0.8$ (red), respectively, where for $\gamma<1$ the disorder potential is no longer exactly restricted to one orbital.}
\end{figure*}

\twocolumngrid
As an example for EPs occuring in a NH Bloch band setting such as the disorder averaged description of our disordered model system, here we focus on the case of a two-banded model with a generic NH Hamiltonian of the form

\begin{align}
H_e(k) = d_0(k) \sigma_0 + \mathbf{d}(k)\cdot \bm{\sigma},
\end{align}
where $\mathbf{d} = \mathbf{d}_R+ i \mathbf{d}_I$ with $\mathbf{d}_R, \mathbf{d}_I \in \mathbb R^3$ and $d_0 \in \mathbb C$. The eigenenergies of  $H_e$ are $E_\pm = d_0 \pm \sqrt{\mathbf{d}_R^2 - \mathbf{d}_I^2 + 2 i \mathbf{d}_R \cdot \mathbf{d}_I}$ and at degeneracy points where the real and the imaginary part under the square root vanish simultaneously, i.e. $\mathbf{d}_R^2 - \mathbf{d}_I^2 = 0$ and $\mathbf{d}_R \cdot \mathbf{d}_I = 0$, the eigenvectors coalesce, leading to an EP. In conclusion, EPs require the tuning of two separate conditions, which is also the case for more than two bands, and thus are a generic and stable phenomenon in systems with at least two spatial dimensions. 

From a mathematical point of view, EPs are the end points of a branch cut in the complex plane, which manifests itself as a line of purely real or purely imaginary energy gap that connects the EPs and corresponds to the conditions  $\mathbf{d}_R \cdot \mathbf{d}_I = 0$ and  $\mathbf{d}_R^2 - \mathbf{d}_I^2 > 0$ or $\mathbf{d}_R^2 - \mathbf{d}_I^2 < 0$ in the two-banded case.  In the context of NH physics, these lines are referred to as (NH) Fermi  arcs and they play a central role in the interpretation of the anomalous transport properties of the model studied in the following.

\subsection{Complex spectrum of the model}
For the system at hand, the simplest possible perturbation, consisting only of an on-site disorder potential, is reflected by the parameter choice $s_0 =1$, $\gamma = 0$, and leads to a trivial self energy $\Sigma$, such that the real part of the spectrum remains the same as for $H_0$ (cf. Fig.~\ref{spec_H0}) and
the imaginary part of the spectrum is flat and twofold degenerate with a value of about $-0.4 i$.

Upon imbalancing the amplitude of the disorder potential between the orbitals, the system enters an exceptional phase. We start by considering the extremal case of the disorder only affecting the $A$-sublattice through the parameter choice $s_0 = 1$, $\gamma = 1$ and $\phi = 0$, noting that deviations from this limiting case will be discussed further below. The Dirac cones are split into EPs connected by a Fermi arc of purely imaginary energy gap (marked by red lines in Fig.~\ref{spec_Heff}), which increases in length with the disorder strength $\alpha$ and is perpendicular to the $k_\mathrm{x}$-direction. By tuning the parameter $\phi$ in Eq.~(\ref{disorder}), the angle between the Fermi arcs and the $k_\mathrm{x}$-axis may be adjusted, where the limiting cases $\phi = 0$ (disorder on the $A$-sublattice) and $\phi = \pi$ (disorder on the $B$-sublattice) afford a simple physical interpretation. Other values of $\phi$ amount to a rotation around the $\sigma_y$ axis, thus resulting in orbital-selectivity in a rotated orbital basis. More generally, the parameter $\phi$ allows us to investigate the influence on the transmission of the angle between the Fermi arcs and the transport direction (cf.~\ref{Transmission_phi_dependent}) in our setup. In an actual experiment, the angle dependence could be readily probed by measuring the transmission in different directions through the sample. 

Another notable feature of the exceptional phase with orbital selective disorder is the presence of states  with a vanishing imaginary part at all energies (marked by green lines in Fig.~\ref{spec_Heff}). As we demonstrate in section~\ref{Sec:Surviving_Bloch_Modes}, these states are Bloch modes of the clean system that remain unaffected by the disorder potential and are responsible for the enhanced and anisotropic transport capabilities of the exceptional phase.

Since the self-energy $\Sigma$ shows no dependence on $\boldsymbol k$ (see Appendix~\ref{perturbation_theory}), the splitting looks similar around all nodal points. It is therefore sufficient to closer inspect the two Fermi arcs in the $(+,+)$ and  $(+,-)$ quadrant and a cut along these two Fermi arcs is shown in Fig.~\ref{spec_Fermi_arc_cut}. The nodal points in the spectrum of $H_0$ from Fig.~\ref{spec_H0} have been inflated into  nodal lines in the real part of the spectrum of $H_e$. By stark contrast to the conventional phase, the DOS at zero energy does not vanish in the quasiparticle picture. To gauge the sensitivity of the exeptional phase towards deviations from the restriction of the disorder potential to one orbital, we also show the result for $\gamma = 0.9$ and $\gamma = 0.8$ in Fig.~\ref{spec_Fermi_arc_cut}, which corresponds to distributing 5\% and 10\% of the disorder amplitude on the second orbital. With decreasing value of $\gamma$, the Fermi arcs shrink and the top of the "Fermi bubbles" starts to move away from zero, i.e. the long-lived states acquire a long but finite life-time.

For all sets of disorder parameters, we find only a very weak if any $\omega$ dependence of $H_e(\bm k, \omega)$, which allows us to draw conclusions on the quasiparticle-dispersion through the complete energy range from $H_e(\bm k, \omega = 0)$ or Fig.~\ref{spec_Heff}, respectively.

\section{Transport signatures} \label{Sec:Transport}
Here, we investigate the influence of the quasiparticle dispersion inside the bulk on the transport properties of a finite-sized sample for the different NH phases. To this end, we generate a random rectangular instance of the system
with $N_\mathrm{x} = 100$  sites in $x$-direction and $N_\mathrm{y} = 200$ sites in $y$-direction. After attaching two spin-independent metallic leads to both ends of the sample, we calculate the two-terminal linear-response conductance 
 in $x$-direction. The dispersion in the leads is given by the Bloch Hamiltonian $H_\mathrm{L} = (\cos(k_\mathrm{x}) + \cos(k_\mathrm{y})) \sigma_0$.

\subsection{Energy-dependent conductance}
First, we present the energy-dependent conductance through the clean system in Fig.~\ref{Transmission_clean}. Resonance effects lead to small fluctuations of the transmission anplitude over the energy spectrum and the vanishing DOS at zero energy completely suppresses transport there.

Fig.~\ref{Transmission_3_in_1} shows the energy-dependent conductance for different disorder configurations. The green line represents the conventional disordered phase, where almost no transport occurs at any energy, which is a consequence of the large imaginary part and therefore damping of modes in the corresponding quasiparticle dispersion. The orange line belongs to  the exceptional disordered phase with $\phi =0$, where the Fermi arcs are perpendicular to the direction of transport along $k_\mathrm{x}$. There, an enhanced conductance at all energies when compared to the conventional phase is visible, carried by the states with an extended lifetime in the vicinity of the green lines in Fig.~\ref{spec_Heff}. Strinkingly, the transmission at zero energy surpasses that of the clean system by far. We interpret that as a signature of the increased DOS in the quasiparticle picture caused by the inflation of the nodal point into a Fermi arc (compare to Fig~\ref{spec_Fermi_arc_cut}). Finally, the blue line indictates the exceptional disordered phase with $\phi =\pi/2$, i.e. with the Fermi arcs parallel to direction of transport. Here, transport is suppressed heavily in comparison to the case with $\phi = 0$.

\subsection{Effect of $\phi$ on the conductance}

\begin{figure*}[htp!]	 
  \centering 
  \sidesubfloat[Conventional phase]{ \includegraphics[trim={1.1cm 0cm -0.1cm 0cm}, width=0.45 \linewidth]{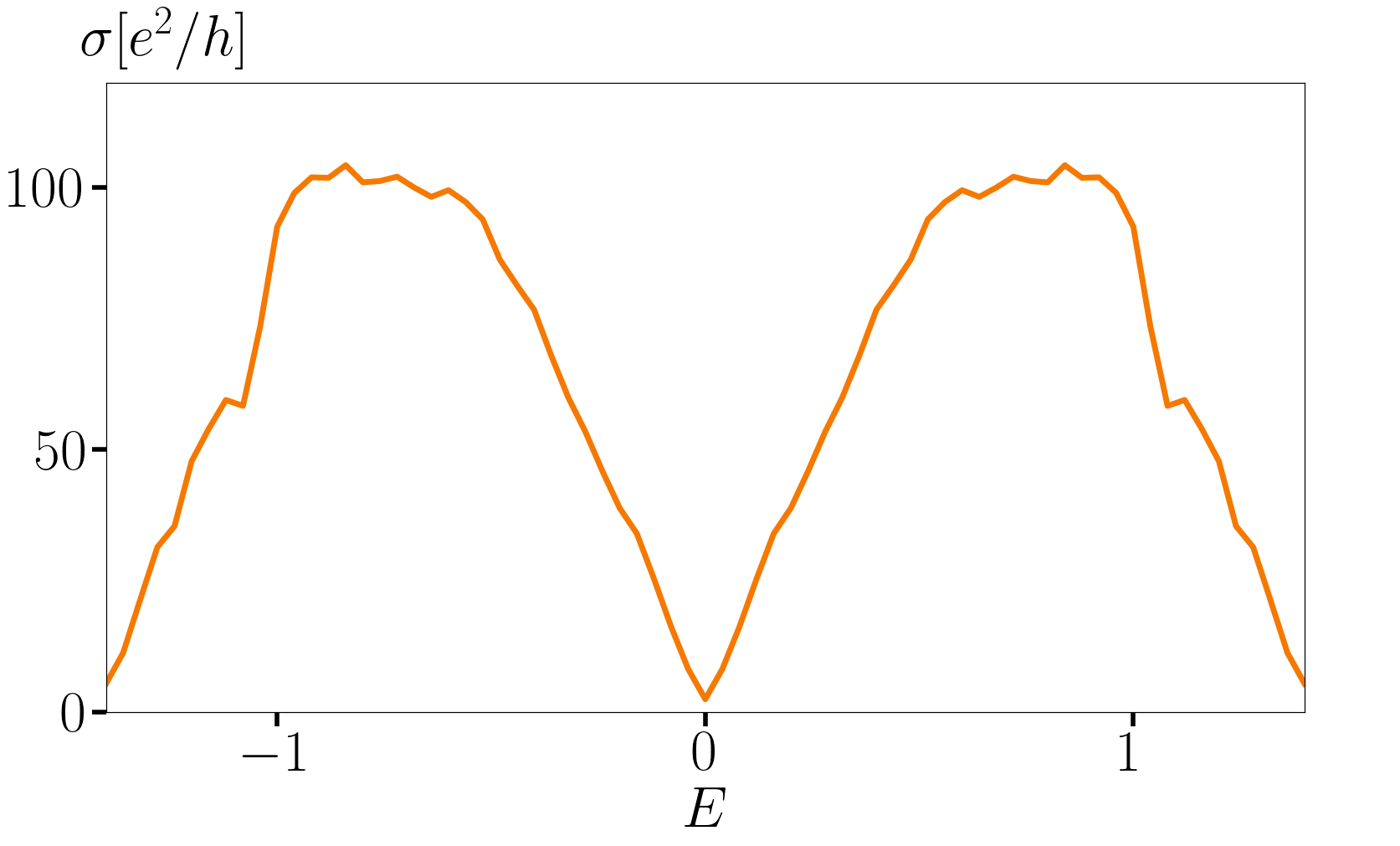}
 \label{Transmission_clean}}
  \sidesubfloat[Clean system]{\includegraphics[trim={0cm 0cm 0cm 0cm},clip, width=0.45 \linewidth]{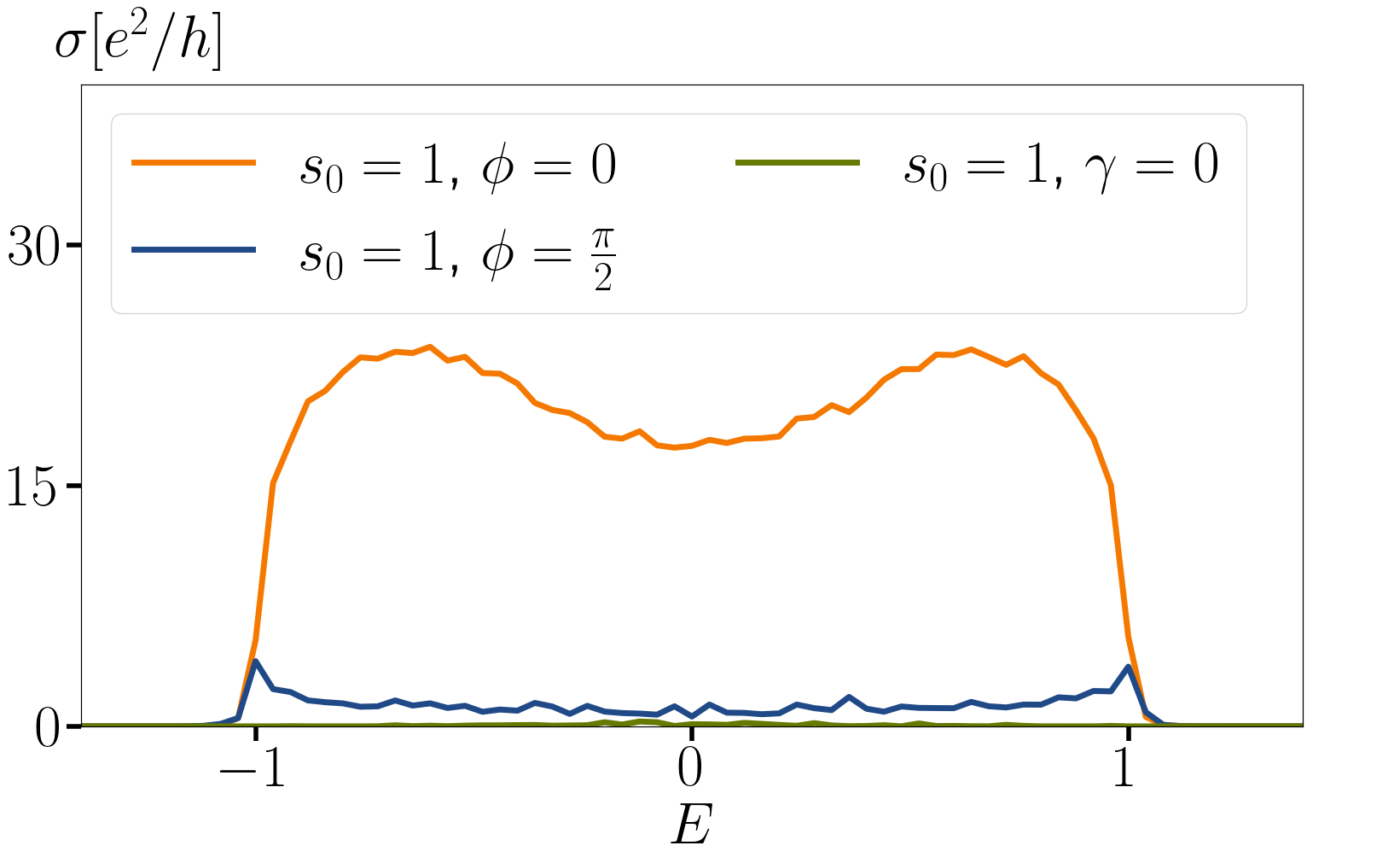}
 \label{Transmission_3_in_1}} \\ 
  \sidesubfloat[Exceptional phase]{\includegraphics[trim={0cm 0cm 0cm 0cm},clip, width=0.45 \linewidth]{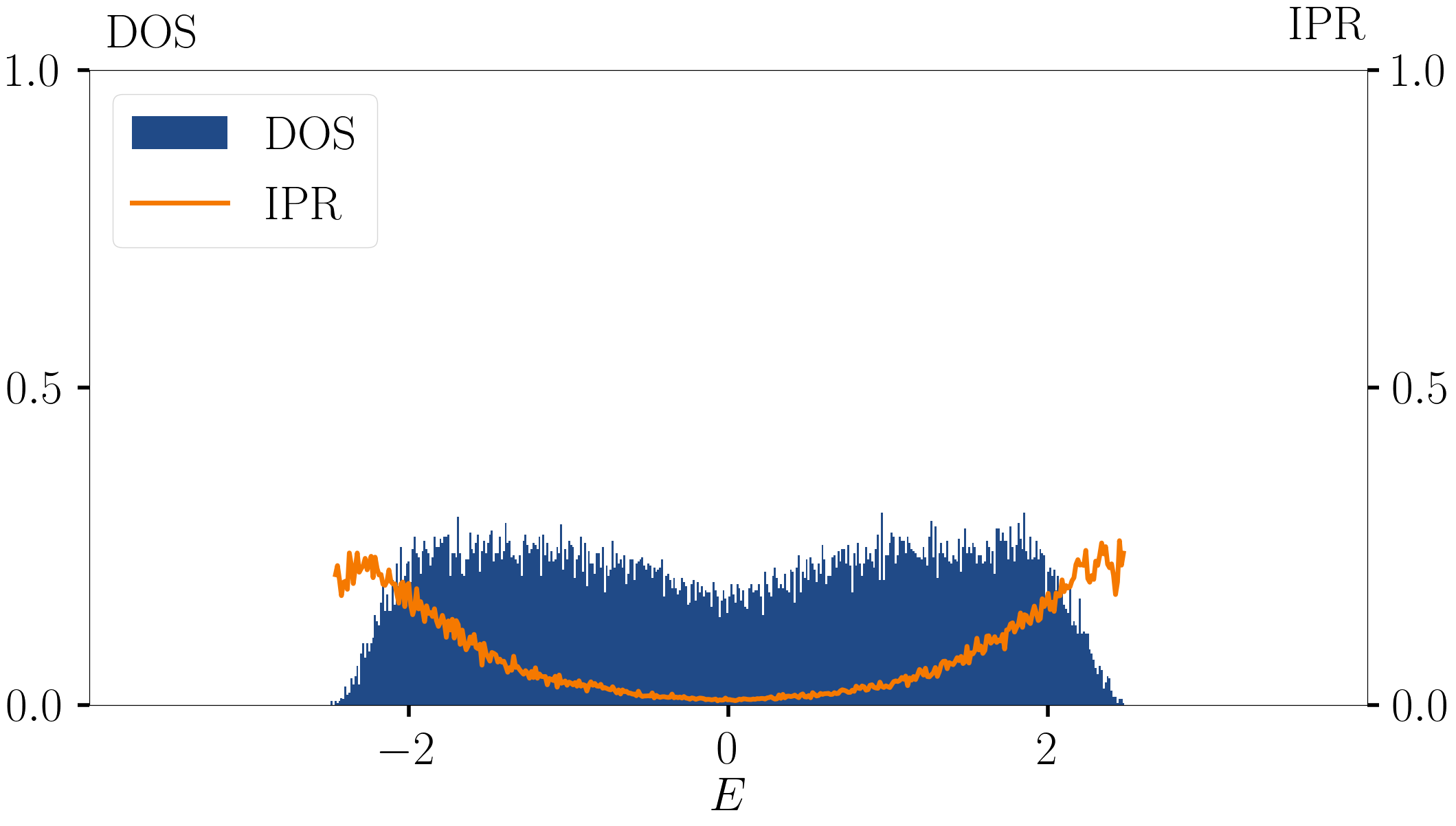}
\label{DOS_IPR_conventional} }
  \sidesubfloat[Exceptional phase]{ \includegraphics[trim={0cm 0cm 0cm 0cm},clip, width=0.45 \linewidth]{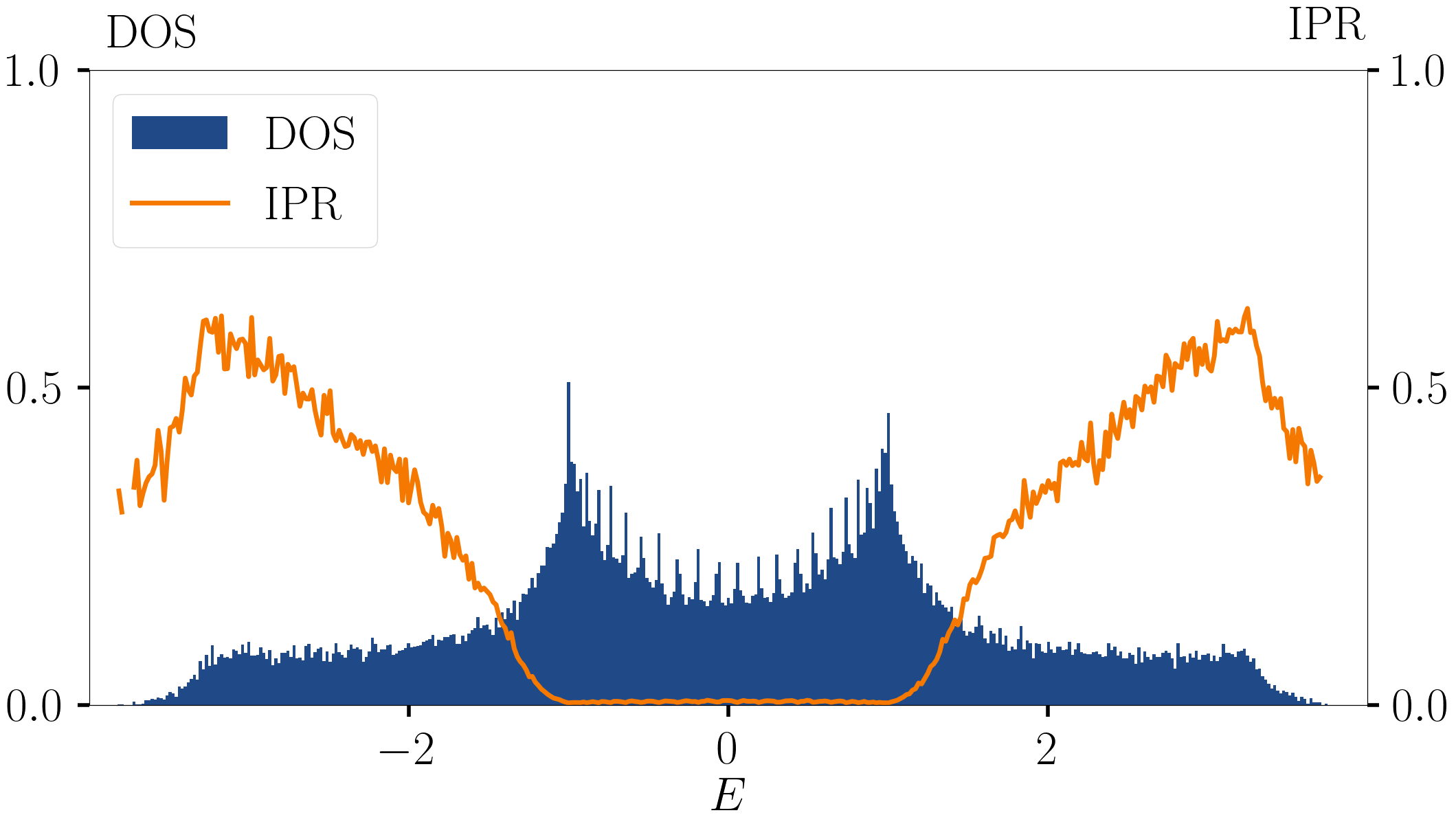} 
 \label{DOS_IPR_exceptional}}
\caption{(a) Energy-dependent two-terminal conductance in $x$-direction for the disorder-free system of size
 $N_\mathrm{x} = 100$, $N_\mathrm{y} = 200$.
 (b) Energy-dependent two-terminal conductance in $x$-direction for various disorder parameters at system size $N_\mathrm{x} = 100$, $N_\mathrm{y} = 200$.
Orange line: exceptional phase with disorder parameters  $s_0 = 1$, $\gamma = 0$, $\phi = 0$, $\alpha = 1.5$. Blue line: exceptional phase with disorder parameters  $s_0 = 1$, $\gamma = 0$, 
$\phi = \frac{\pi}{2}$, $\alpha = 1.5$.  Green line: conventional phase with disorder parameters  $s_0 = 1$, $\gamma = 0$, $\alpha = 1.5$. 
(c) Normalized DOS and average IPR at each energy in the conventional phase. Disorder parameters are $s_0 = 1$, $\gamma = 0$, $\alpha = 1.5$. Data from exact diagonalization
of a system with $N_\mathrm{x} = 50$, $N_\mathrm{y} = 50$ sites. (d) Normalized DOS and average IPR at each energy in the exceptional phase. Disorder parameters are $s_0 = 1$, $\gamma = 1$, $\phi = 0$, $\alpha = 1.5$. Data from exact diagonalization of a system with $N_\mathrm{x} = 50$, $N_\mathrm{y} = 50$. All disordered data averaged over 10 independently drawn realizations. 
 }
\end{figure*}

As Fig.~\ref{Transmission_3_in_1} shows, the angle of the Fermi arc relative to the direction of transport affects transport at all energies in the exceptional phase. More specifically, the transmission decreases with increasing $\phi$, which can be intuitively understood from the quasiparticle dispersion. In the imaginary part of the spectrum for $\phi = 0$  in Fig.~\ref{spec_Heff}, ridges with vanishing imaginary part are visible and accentuated by green lines.
Due to their prolonged or even infinite (at $\gamma=1$) lifetime, it is reasonable to expect that these quasiparticles and the ones close to them mostly carry the transport. Additionally, all of them have a group velocity parallel to the $k_\mathrm{x}$-axis, i.e. the direction of transport. With increasing $\phi$, the Fermi arc and also the ridges rotate in the $k_\mathrm{x}$-$k_\mathrm{y}$-plane, while the direction of transport remains along the $k_\mathrm{x}$-axis. The group velocity of the states associated with the ridges rotates as well and always encloses the angle $\phi$ with the direction of transport. An elementary trigonometric construction shows that the wave fronts will have to travel a longer distance of $N_\mathrm{x} \sqrt{1 + \tan^2(\phi)}$ through the sample (see Appendix~\ref{travelling_distance}), which causes the decay of the transmission amplitude.

 \floatsetup[figure]{style=plain,subcapbesideposition=top} 

\begin{figure}[htp!]	 
  \centering 
 \includegraphics[trim={0cm 0cm 0cm 0cm},clip, width= \linewidth]{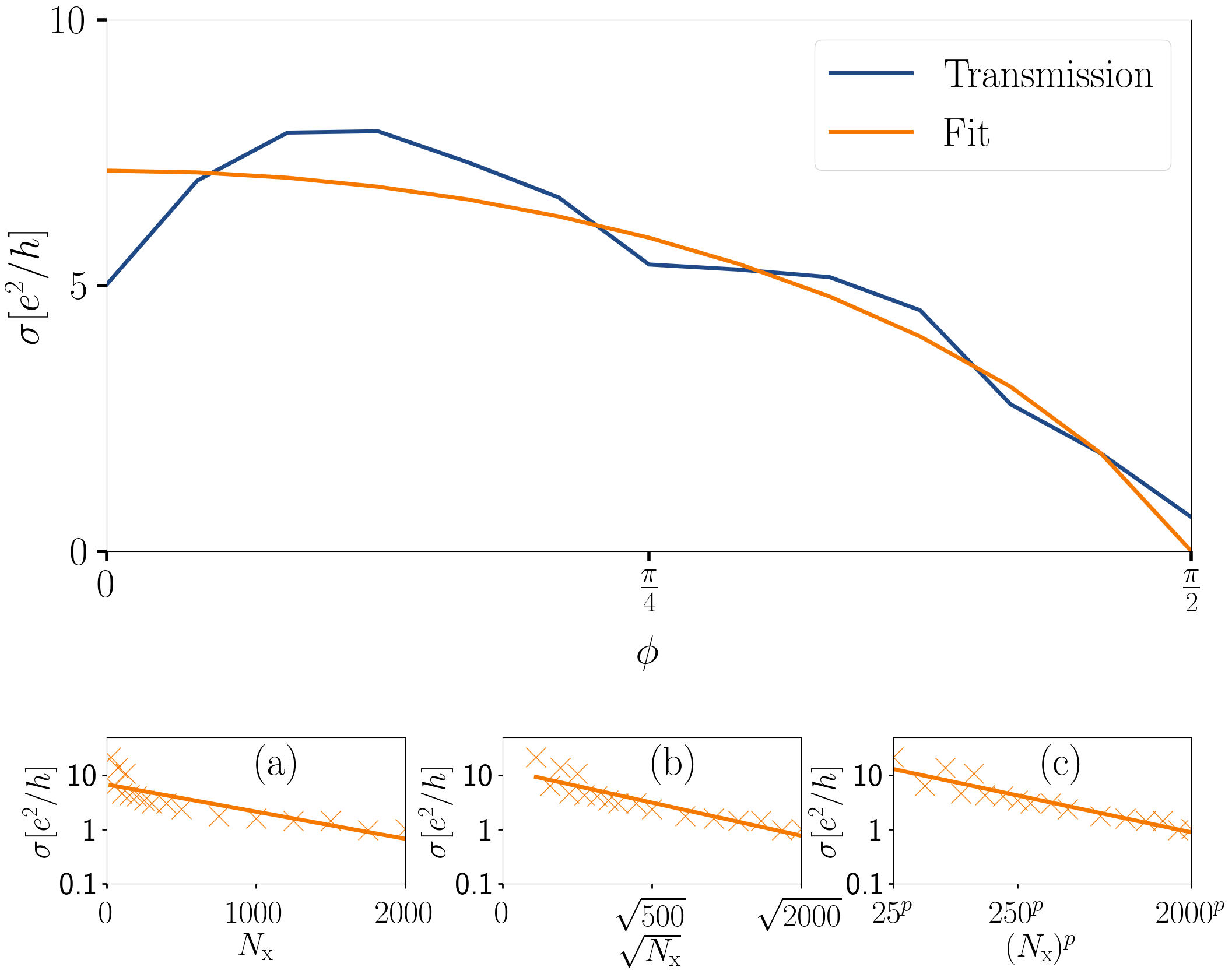} 
\caption{Conductance in $x$-direction at zero energy in the exceptional phase as a function of $\phi$ along with a fit of 
$T(\phi) = \frac{e^2}{h}  \exp \left [ \widetilde{A} -\widetilde{\tau} \left ( N_\mathrm{x} \sqrt{1 + \tan^2(\phi)}  \right ) ^ p \right ]$, where $p= \frac{1}{5}$, with the parameters $\widetilde{A} = 4.67$, $\widetilde{\tau} = 1.08$. Remaining disorder parameters are $s_0 = 1$, $\gamma = 1$, $\alpha = 1.5$. System size $N_\mathrm{x} = 100$, $N_\mathrm{y} = 200$. Insets show the decay of the conductance at zero energy for increasing system size along with a fit of $T(l) = \frac{e^2}{h} \exp \left [A - \tau \left (  N_\mathrm{x} \right ) ^ p \right ]$. Remaining disorder parameters are $s_0 = 1$, $\gamma = 1$, $\phi = 0$,  $\alpha = 1.5$. System $N_\mathrm{y} = 200$.  
	 (a) $p = 1$ and $A  = 1.92$, $\tau = 0.0012$. (b) $p= \frac{1}{2}$ and $A  = 2.61$, $\tau = 0.068$. (c) $p= \frac{1}{5}$ and $A  = 4.63$, $\tau = 1.06$. 
\label{Transmission_phi_dependent}}
\end{figure}

To further substantiate this influence of the parameter $\phi$ on the transmission, we present the conductance in the exceptional phase at zero energy in dependence on $\phi$ in the main plot of Fig.~\ref{Transmission_phi_dependent} for a fixed system length of $N_\mathrm{x} = 100$ sites, along with the dependence on system length in the insets (a) - (c). In order to mitigate the influence of the coupling to the leads and varying degrees of reflection with $\phi$, we work close to the wide-band limit by flattening the dispersion in the leads to  $H_\mathrm{L} = 0.1 (\cos(k_\mathrm{x}) + \cos(k_\mathrm{y})) \sigma_0$ for the calculations in Fig.~\ref{Transmission_phi_dependent}.

We start by discussing the insets (a), (b) and (c) of Fig.~\ref{Transmission_phi_dependent}, which show the decay of the transmission in dependence of system length  for $\phi = 0$ along with a fit of the law $T(l) = \frac{e^2}{h} \exp \left [A - \tau \left (  N_\mathrm{x} \right ) ^ p \right ]$ for decreasing values of $p$. Specifically, we have $p= 1$ in (a), $p= \frac{1}{2}$ in (b), and $p= \frac{1}{5}$ in (c). Apart from small resonance-induced fluctuations at short system lengths, the fit in inset (c) with $p= \frac{1}{5}$ is closest to the actual data and yields the parameters  $A  = 4.63$, $\tau = 1.06$.

This fit to the length-dependent transmission is motivated by previous work on disordered one-dimensional systems suggesting that such a slower-than-exponential decay of the transmission amplitude is tied to anomalous localization properties \cite{anom_loc1, anom_loc2, anom_loc3}. There, the value of $p$ that determines the length dependent transmission decay corresponds to 
the value obtained from the localization of the wave-function as $\Psi(x) \propto \exp  \left [- \lambda |x|^p \right ]$, where $p = 1$  for standard Anderson localization.  Since our fits agree best with the if we choose $p= \frac{1}{5}$ or smaller, it seems plausible that the exceptional phase features very weakly localized microscopic eigentstates, which is also indicated by a more careful analysis of the localization properties in section~\ref{Sec:Localization}.

Moving to the main plot of Fig.~\ref{Transmission_phi_dependent}, a decay of the conductance with $\phi$ is clearly visible. As mentioned before, the distance that the modes travel through the sample increases with $\phi$ as $N_\mathrm{x} \sqrt{1 + \tan^2(\phi)}$. Taking into account the findings from the insets, the $\phi$-dependent conductance  should roughly follow a law of the form  $T(\phi) = \frac{e^2}{h}  \exp \left [ \widetilde{A} -\widetilde{\tau} \left ( N_\mathrm{x} \sqrt{1 + \tan^2(\phi)}  \right ) ^ p \right ]$ with $p= \frac{1}{5}$. A fit reveals a good agreement and basically recovers the parameters from the fit to the length-dependent decay with $\widetilde{A}  = 4.67$, $\widetilde{\tau} = 1.08$.

\subsection{Effect of the disorder amplitude on the conductance}

In  Fig.~\ref{Transmission_length_arc}, we analyze the dependence of the conductance in the exceptional phase on the disorder strength $\alpha$.  The parameters $s_0 = 1$ and $\phi = 0$ are fixed and we explore the behaviour for  the ``ideal'' case $\gamma = 1$ with infinitely lived quasiparticles as well as $\gamma = 0.9$, and $\gamma = 0.8$, where the disorder potential is no longer restricted to precisely one orbital such that the Fermi bubbles no longer touch zero (cf.~Fig.~\ref{spec_Fermi_arc_cut}). Quite remarkably, both for $\gamma = 1$ and $\gamma=0.9$, the conductance increases with disorder strength, together with the length of the Fermi arc which is plotted alongside. For $\gamma = 0.8$, the conductance exhibits a similar tendency at small disorder amplitudes, but drops again for strong disorder, where the increasing overall damping exceeds the inflation of the Fermi arcs.

The conductance enhancing effect of disorder in the exceptional phase is a consequence of the growing quasiparticle DOS and size of the imaginary Fermi bubbles (cf. Fig.~\ref{spec_Fermi_arc_cut}), which increases the amount of states with prolonged life-time (suppressed damping). In this sense, the quite counter-intuitive observation of a conductance that increases with disorder strength affords a simple explanation within the framework of our effective NH Hamiltonian analysis. Our transport simulations are performed using the Kwant library \cite{Kwant}.

\section{Microscopic localization properties}\label{Sec:Localization}

Here, we turn to the microscopic properties of the system. We calculate the inverse participation ratio (IPR) and the actual DOS obtained from a full diagonalization of the microscopic Hamiltonian. The IPR is a measure for the localization of states, which is given by 
$\int \mathrm{d}\bm r |\Psi(\bm r)|^4$. 

\subsection{Microscopic DOS and average IPR}
Fig.~\ref{DOS_IPR_conventional} shows the microscopic DOS and the averaged IPR at each energy for the disordered system in the conventional phase with the disorder parameters set to $s_0 =1$, $\gamma = 0$, $\alpha = 1.5$. The localization strength varies with energy and  is weakest
at zero energy with an IPR of about $0.008$. It continuously increases when moving away from this point. The DOS at zero energy does not vanish (in contrast to the clean system) as a consequence of the $\sigma_0$-term in the disorder (cf. Eq.~(\ref{disorder})) that breaks the chiral symmetry of $H_0$. 

For the exceptional phase in Fig.~\ref{DOS_IPR_exceptional} with the disorder parameters set to $s_0 =1$, $\gamma = 1$, $\phi =0$, $\alpha = 1.5$, the states are  uniformly localized for energies between -1 and 1, with an IPR of about $0.007$. Outside of that range, the localization grows stronger rapidly. The interval of relatively weak localization in the exceptional phase roughly coincides with the energy range in which transmission occurs (c.f. Fig.~\ref{Transmission_3_in_1}) and also with the range that is covered by the real part of the spectrum (c.f. Fig.~\ref{spec_Heff}). Although the microscopic DOS covers a larger energy interval than in the conventional phase, no qualitative difference is visible.

\subsection{Surviving Bloch modes in the exceptional phase}
\label{Sec:Surviving_Bloch_Modes}

\begin{figure*}[htp!]	 
  \centering 
  \sidesubfloat[Conventional phase]{ \includegraphics[trim={0cm 0cm 0.cm 0cm}, width=0.5 \linewidth]{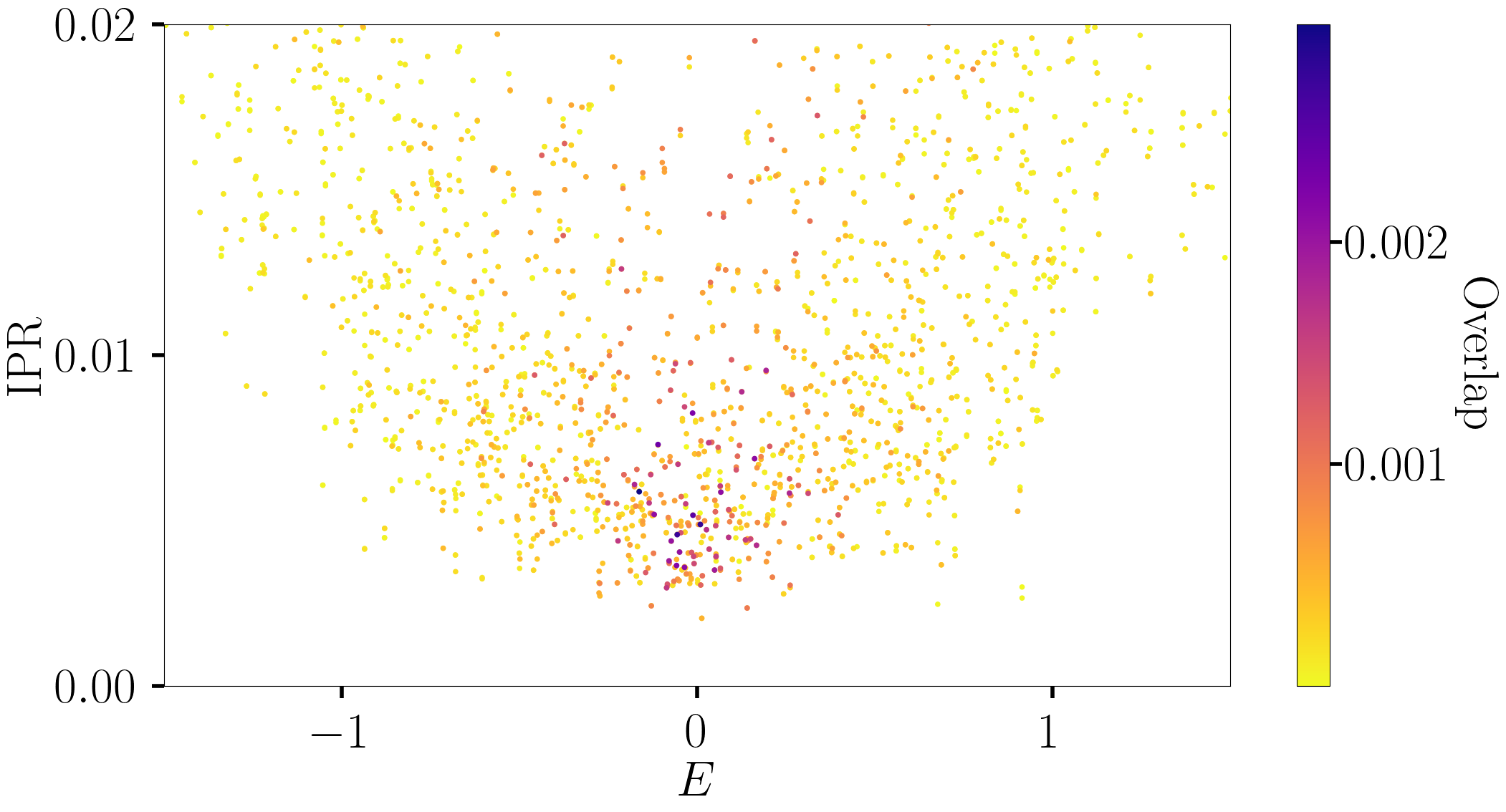}
 \label{overlap_conventional}}
  \sidesubfloat[Clean system]{\includegraphics[trim={0cm 0cm 0cm 0.cm},clip, width=0.49 \linewidth]{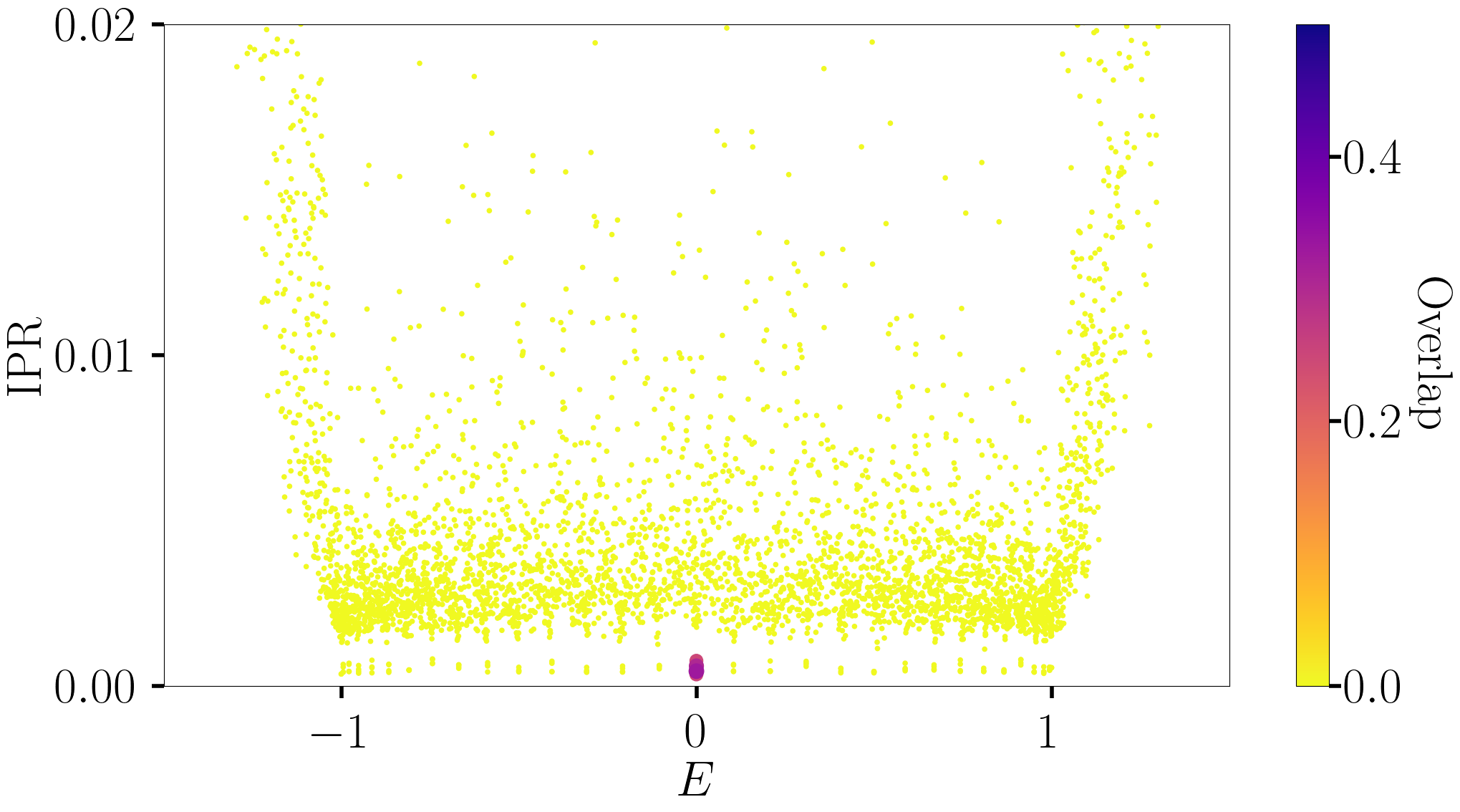}
\label{overlap_exceptional}
 }
\caption{Scatter plot of the microscopic Hamiltonian eigenstates over energy and IPR for a randomly generated system of $60\times60$ sites. The point size and color indicate the overlap $|\langle \boldsymbol k | ES \rangle |^2$ of the eigenstate $|ES\rangle$ with the  eigenstate $|\boldsymbol k \rangle$ of the effective Hamiltonian $H_\mathrm{e}$ at $k_\mathrm{x} = \pi/2$, $k_\mathrm{y} = \pi/2$ with the smaller imaginary part, i.e. lower damping. (a) Result for a system in the conventional phase with parameters $s_0 = 1$, $\gamma = 0$, $\alpha = 1.5$.  
(b) Result for a system in the exceptional phase with parameters $s_0 = 1$, $\gamma = 1$, $\phi = 0$, $\alpha = 1.5$.  Note that the plot range only covers the energies for which transport occurs and also cuts off some of the more strongly localized states with very high IPRs. \label{overlaps}
 }
\end{figure*}

While the average IPR around zero energy is only marginally smaller in the exceptional phase (cf. Fig~\ref{DOS_IPR_conventional},~\ref{DOS_IPR_exceptional}), we find a clear distinction in the distribution of the individual IPRs of the microscopic eigenstates between the two phases. Fig.~\ref{overlaps} compares the individual IPR values in both phases as a scatter plot over the energies of their respective eigenstates. The conventional phase in~\ref{overlap_conventional} features eigenstates with varying degree of localization such that the minimum value of the IPR at zero energy  is about 0.003 or higher. By contrast, the exceptional phase depicted in ~\ref{overlap_exceptional} possesses a set of completely delocalized states that come in clusters of four throughout the energy range from -1 to 1.  

These delocalized states are Bloch waves of the clean system that live entirely on the sublattice unaffected by the random disorder and thus survive under any perturbation amplitude. This phenomenon is not a fine-tuned peculiarity of our model but rather a manifestation of a more general {\em theorem}: 

{\em Consider an arbitrary tight-binding model $H_0$ with translational invariance and a number of $n$ orbitals. For a random disorder term $V$ that only affects a subset of $n_\alpha < n$ orbitals, there can be Bloch modes of $H_0$ that live on the remaining $n - n_\alpha$ orbitals and thus are still eigenstates of the disordered system $H = H_0 + V$ regardless of the concrete nature of $V$. Such modes require the tuning of $n_\alpha$ complex conditions in momentum space, which can be made real or reduced in number by the presence of certain symmetries (see Appendix~\ref{APP:Bloch_States} for a proof).}

These insights complement the analysis of Ref.~\cite{anom_loc4}, where the existence of zero-energy (localized) eigenstates that are restriceted to one sublattice of a disordered bipartite lattice has been derived.

The surviving Bloch modes make up the quasiparticles with vanishing imaginary energy that we observe in the spectrum of the effective Hamiltonian (cf. Fig.~\ref{spec_Heff}). To illustrate this, we represent the inifinitely-lived eigenstate of $H_\mathrm{e}$ with momentum $k_\mathrm{x} = \pi/2$, $k_\mathrm{y} = \pi/2$ (in the center of the Fermi arc) in the basis of the microscopic eigenstates of the system and visualize the result in Fig.~\ref{overlap_exceptional} by the color and size of the plot points. The quasiparticle is comprised entirely of the  four delocalized Bloch eigenstates from the cluster at zero energy. The other three eigenstates of $H_\mathrm{e}$ with zero (real) energy and infinite lifetime belonging to the other three Fermi arcs can also be represented within the same cluster. A similar outcome is observed for the non-decaying quasiparticles at other energies.

On the other hand, the quasiparticles with a large imaginary energy part are composed of a wide range of microscopic states with high IPR-values. Exemplarily, we show the overlap of one of the quasiparticles in the conventional phase (where all energies  have an imaginary part of about $-0.4i$) with the microscopic eigenstates of the system in Fig.~\ref{overlap_conventional}.

\begin{figure}[tp]	 
  \centering 
 \includegraphics[trim={0cm 0cm 0cm 0cm},clip, width= \linewidth]{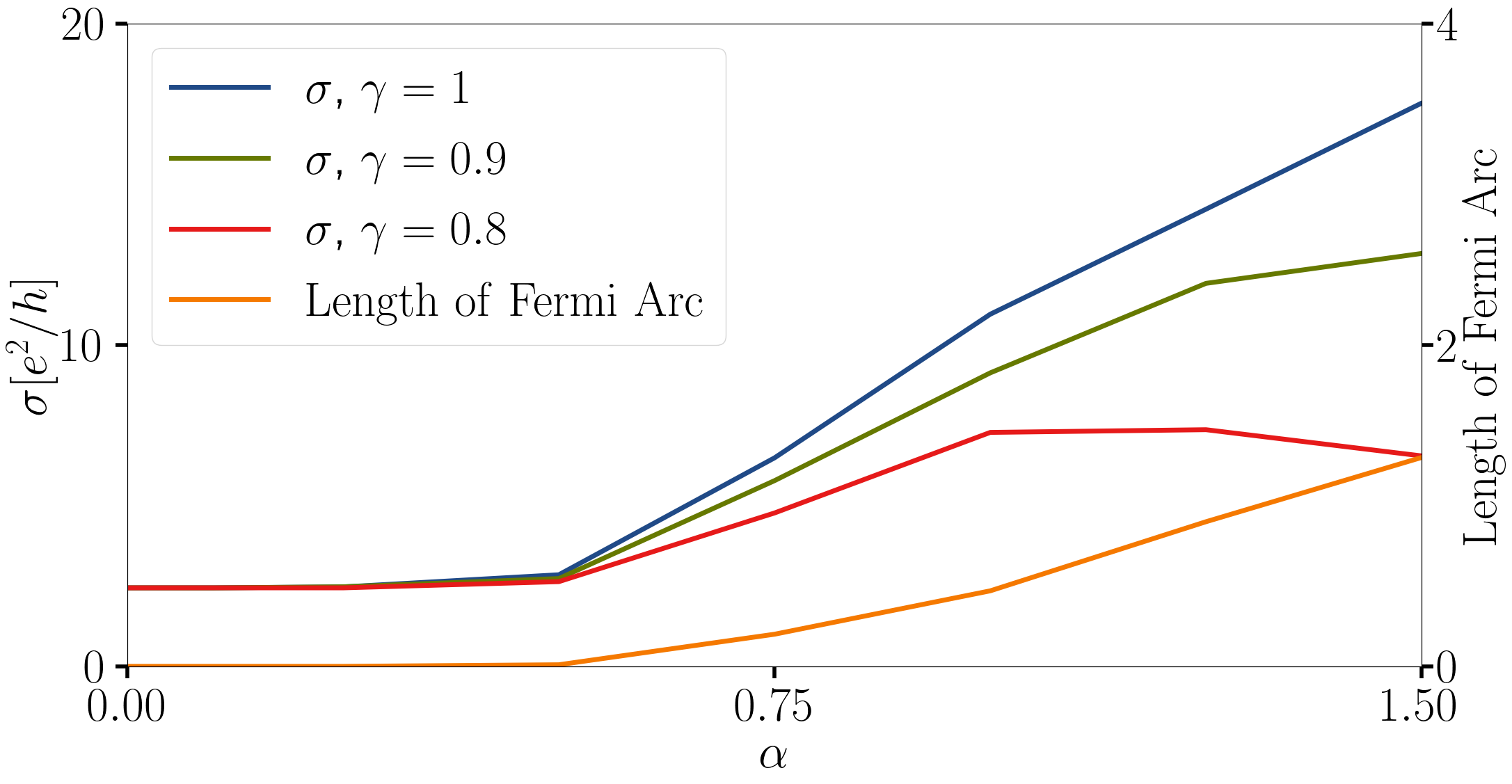} 
\caption{Conductance  in $x$-direction at zero energy in the exceptional phase as a function of disorder strength $\alpha$. 
Disorder parameters are $s_0 = 1$, $\phi = 0$ for all curves, while we vary $\gamma = 1$, $0.9$ and $0.8$ to study deviations from perfectly orbital-selective disorder ($\gamma=1$). System size is $N_\mathrm{x} = 100$, $N_\mathrm{y} = 200$. For comparison, the length of the Fermi arc is shown for $s_0 = 1$, $\phi = 0$, $\gamma = 1$.
 \label{Transmission_length_arc} }
\end{figure}

 \floatsetup[figure]{style=plain,subcapbesideposition=top} 

The symmetry of our model system reduces the one complex condition, that arises from a disorder term restricted to one orbital, to a real one (see Appendix~\ref{APP:Bloch_States}). In conclusion, the surviving Bloch states have codimension one and should form one-dimensional submanifolds in the momentum space of our two-dimensional system, which are precisely the green lines in Fig.~\ref{spec_Heff}. 

Even though there is no direct connection between the surviving Bloch modes and the occurrence of EPs in the spectrum of the effective Hamiltonian, the two effects play together to create the observed transport phenomenology. While the Bloch states themselves only provide a single transport channel for a given energy regardless of the system size, they enforce the lines of vanishing imaginary energy in the spectrum of the effective Hamiltonian (cf. Fig.~\ref{spec_Heff}) and thereby a population of quasiparticles with a very long lifetime in the immediate vicinity of these lines. The transmission amplitude carried by these regions of the spectrum does also scale with system size. Since the distance between the EPs grows with disorder strength, the associated Fermi bubbles broaden and bring the imaginary part of the spectrum surrounding the Bloch states closer to zero (cf. Fig~\ref{spec_Fermi_arc_cut}), thereby increasing the transmission amplitude (cf. Fig.~\ref{Transmission_length_arc}).

\subsection{Localization behaviour}
Based on the methods used in an earlier study on anomalous localization \cite{anom_loc4}, we measure the spatial decay of the probability amplitude relative to the maximum by defining the correlation function 

\begin{align}
g(r, \varphi) = \left \langle \log \left( \frac{|\Psi_\mathrm{max}|^2}{|\Psi(r, \varphi)|^2} \right) \right  \rangle. \label{correlation_function}
\end{align}
Here, we consider the probability density summed over the two internal degrees of freedom, i.e. $|\Psi|^2 = |\Psi_A|^2 + |\Psi_B|^2$, let $|\Psi_\mathrm{max}|^2$ denote the maximum, $|\Psi(r, \varphi)|^2$ the probability density in relative polar coordinates to the location of the maximum, and $\langle ... \rangle$ the average over multiple eigenstates. For an average localization behaviour $\Psi(r, \varphi) \propto \exp  \left [- \lambda r^{p(\varphi)} \right ]$ with $0 < p(\varphi) < 1$), the correlation function should behave as $g(r, \varphi) = \lambda r^{p(\varphi)}$.

In general, our data agrees with the above ansatz and suggests that there is no dependence of the value of $p$ on $\varphi$ for larger systems with similar dimension in $x$ and $y$ direction. To properly capture the asymptotic localization behaviour, we investigate systems with a size of $N_\mathrm{x} = 5000$, $N_\mathrm{y} = 200$ by using the Krylov solver of the sparse matrix library of Scipy \cite{Scipy} to obtain 100 eigenstates close to zero energy of a randomly generated instance of the system, where we exlude the surviving Bloch eigenstates in the exceptional phase. From those, we compute the correlation function as per Eq.~(\ref{correlation_function}) in the direction $\varphi = 0$, i.e. along the $x$ direction. 

\begin{figure*}[htp!]	 
  \centering 
  \sidesubfloat[Conventional phase]{ \includegraphics[trim={0cm 0cm 0.cm 0cm}, width=0.47 \linewidth]{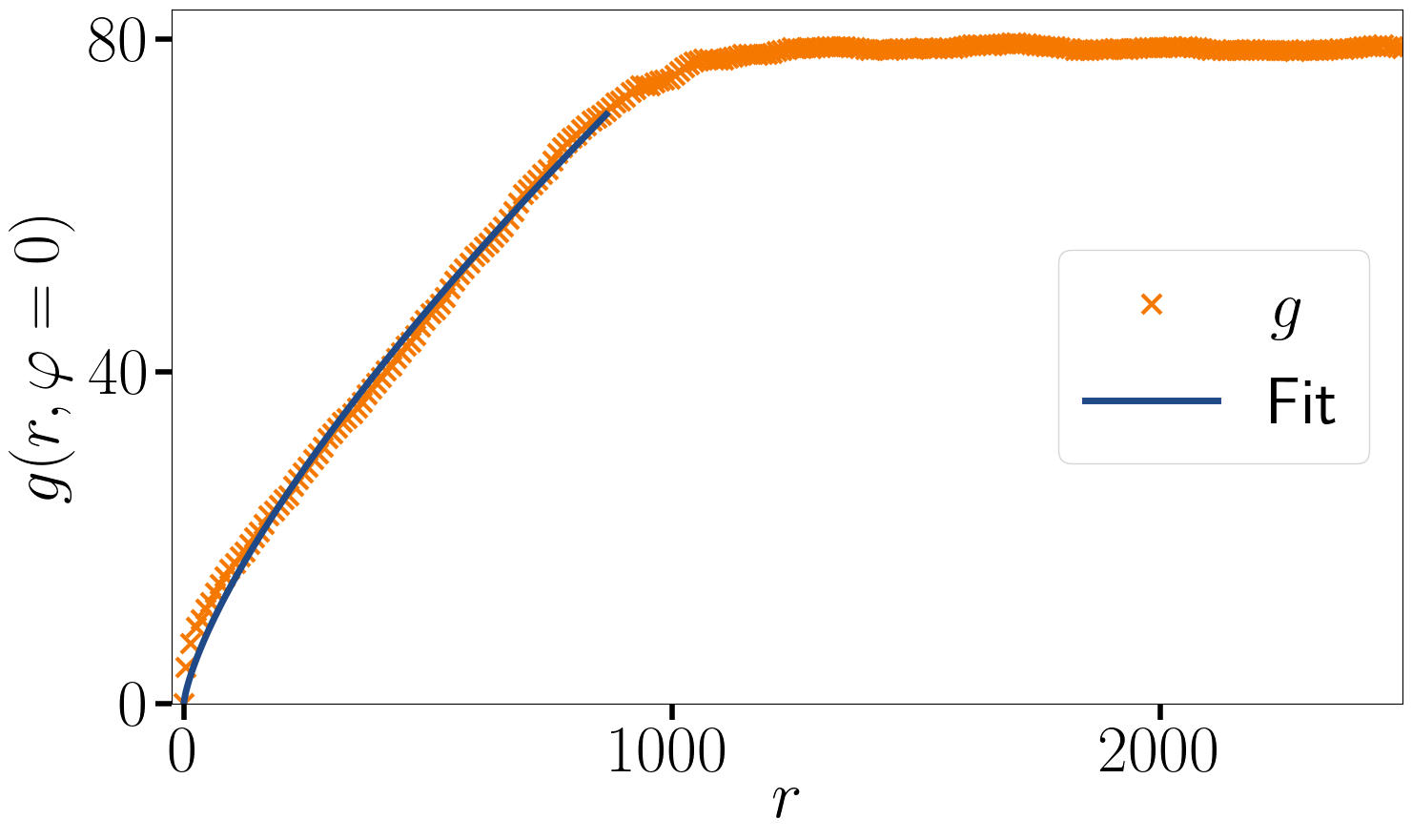}
 \label{CF_conventional}}
  \sidesubfloat[Clean system]{\includegraphics[trim={0cm 0cm 0cm 0cm},clip, width=0.47\linewidth]{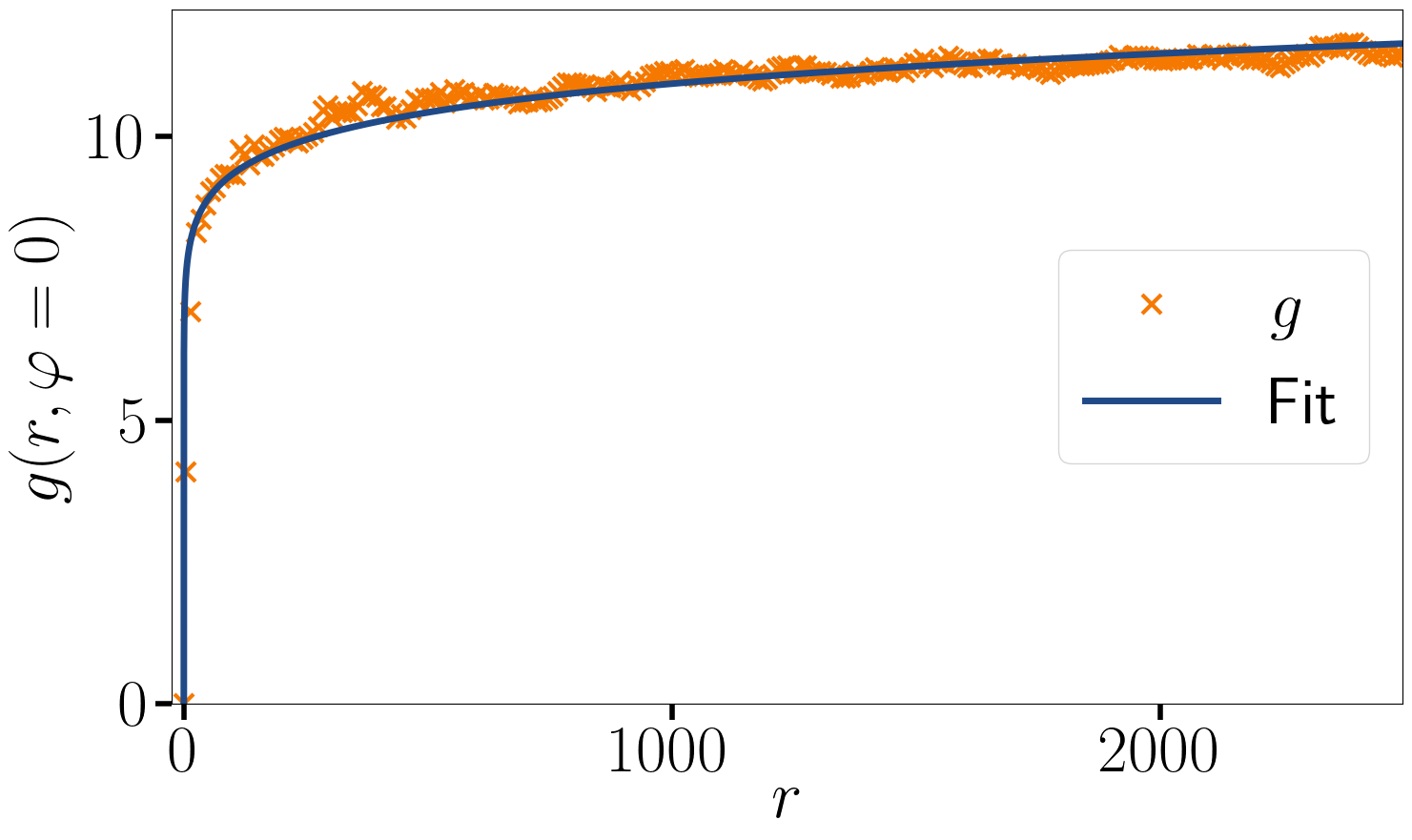}
\label{CF_exceptional}
 }
\caption{The correlation function $g(r, \varphi = 0)$ (see Eq.~(\ref{correlation_function})) for system size $N_\mathrm{x} = 5000$, $N_\mathrm{y} = 200$ along with 
a fit of the form $g(r, \varphi = 0) = \lambda r^p$ . (a) Result for the conventional phase with parameters $s_0 = 1$, $\gamma = 0$, $\alpha = 1.5$. Fit parameters are $\lambda = 0.5$, $p = 0.73$, where only the data up to the saturation resulting from the limited machine precision was used. (b) Result for the exceptional phase with parameters $s_0 = 1$, $\gamma = 1$, $\phi = 0$, $\alpha = 1.5$. Fit parameters are $\lambda = 6.82$, $p = 0.07$.}
\end{figure*}

Fig.~\ref{CF_conventional} shows $g(r, \varphi = 0)$ for the conventional phase, which behaves roughly linear up to the point where $\Psi(r, \varphi = 0)$ becomes smaller than machine precision and the correlation function saturates. A fit of the form $g(r, \varphi = 0) = \lambda r^p$ yields $\lambda = 0.5$, $p = 0.73$. For other sample geometries and perturbation amplitudes $\alpha$, given that the length of the sample in the direction of $\varphi$ is big enough, we observe similar behaviour and obtain fit parameters $p$ ranging between $0.7$ and $1$, which we deem compatible with regular Anderson localization.

For the exceptional phase in Fig.~\ref{CF_exceptional}, a much slower and non-linear decay of the correlation function is observed. At the maximum system length presented here, $g(r, \varphi = 0)$ is far from saturating and the fit $g(r, \varphi = 0) = \lambda r^p$ reveals the parameters $\lambda = 6.82$, $p = 0.07$. Again, we observe similar behaviour for other sample geometries and obtain values of $p$ around 0.1 from the corresponding fits. In summary, our data provides numerical evidence that the exceptional phase of our model exhibits  anomalous localization. This is also consistent with the sub-exponentially decaying transport amplitude (cf. Fig.~\ref{Transmission_phi_dependent}).

 \floatsetup[figure]{style=plain,subcapbesideposition=top} 

\section{Candidates for Experimental Platforms}
\label{Sec:Platforms}
\subsection{Basic requirements}
Besides the intriguing algebraic aspects of EPs by themselves, a core ingredient for the physics discussed in this work is the possibility of quasiparticles with infinite lifetime in a dissipative environment caused by disorder scattering. Such extraordinarily long-lived modes are associated with Fermi-arc regions in between the EPs, where the imaginary part of the complex spectrum of $H_e$ (at least approximately) touches the real axis (cf. Fig.~\ref{spec_Heff}). For this to occur, we have identified three crucial ingredients.
\newline $(i)$ First, a chiral-symmetric base model $H_0$, which has symmetry-protected nodal points. The symmetry implies that the Bloch Hamiltonian $H_0(\boldsymbol k )$ contains only two of the three Pauli matrices $\sigma_\mathrm{x,y,z}$ (or two linearily independent combinations of them) and reduces the complex condition from the theorem in Section~\ref{Sec:Surviving_Bloch_Modes} to a real one and thus ensures the survival of Bloch modes at all energies.
\newline $(ii)$ Second, a random disorder with on-site terms of the form

\begin{align}
V _\mathrm{\boldsymbol j, \boldsymbol j} \sim a_\mathrm{\boldsymbol j} (s_0 \sigma_0 + \boldsymbol s  \cdot \boldsymbol \sigma),
\label{eqn:disorderingredients}
\end{align}
with $|s_0| = |\boldsymbol s|$.  The vector $\boldsymbol s $ must be chosen such that $\boldsymbol s  \cdot \boldsymbol \sigma$ respects the chiral symmetry of the base model $H_0$. This ensures that the total model can be brought back to a form where the disorder potential only affects one orbital and the Bloch Hamiltonian possesses eigenstates that live on the opposite orbital by a unitary transformation.
\newline $(iii)$ Third,  a symmetric distribution of the random 
amplitudes $\{ a_\mathrm{\boldsymbol j} \}$ around zero is required. Any deviation from a symmetric distribution reduces the length of the Fermi arcs and shifts them away from zero energy.

\subsection{Square-net materials}
As a first candidate, we consider a setup where the two internal degrees of freedom are given by different atomic orbitals. We may choose $\boldsymbol s  = (0,0,1)$ to represent disorder on only one of the two orbitals, since

\begin{equation}
\frac{\sigma_0 +  \sigma_\mathrm{z}}{2} = 
\begin{pmatrix}
1 && 0\\
0 && 0
\end{pmatrix}. \label{orbital_disorder}
\end{equation}
A model of this type may occur within the class of multi-atomic crystals with square sub-nets which have been found to exhibit a Dirac 
dispersion  \cite{SQN_Dirac_1, SQN_Dirac_2}. A remaining challenge then is to restrict the disorder potential (at least to good approximation) to one of the two orbitals of the atomic species that forms the square net. In many cases though, the orbitals are $p$-like as in  LaCuSb$_2$, where the Sb atoms form the square net that yields Dirac physics \cite{SQN_Dirac_3}. Assuming a crystal growth direction that results in a square parallel to the surface, positive and negative ions adsorbed to the surface may affect one of the orbitals much stronger, since they are club-shaped in different spatial directions, thus creating a disorder term similar to Eq.~(\ref{orbital_disorder}). 

Our calculations show that the disorder does not have to be restricted to one orbital perfectly. In particular, if the disorder strength on one of the orbitals is about 5 \% of the other one, the quasiparticles along the green lines in Fig.~\ref{spec_Heff} obtain a small imaginary part, but the characteristic transport signatures still remain (cf.~Fig.~\ref{spec_Fermi_arc_cut} and~\ref{Transmission_length_arc}). Only at about 10 \% deviation from the ideal configuration, the characteristic signatures start to become unrecognizable.

According to the above ingredient $(ii)$, the Bloch Hamiltonian $H_0(\boldsymbol k)$ of the base model should posses a chiral symmetry that allows $\sigma_z$ and one other Pauli matrix. However, this should not be problematic, since the setup with atomic orbitals exhibits real hoppings. Together with the square-shaped lattice, a chiral symmetry that rules out $\sigma_\mathrm{y}$ and permits $\sigma_\mathrm{x}$, $\sigma_\mathrm{z}$ is to be expected. 

Along the same lines of reasoning, honeycomb materials such as graphene or silicene as well as other Dirac materials with non-rectangular lattices do not represent promising candidates to create an exceptional phase akin to the one discussed in this work. The chiral symmetry of such lattices will generally permit $\sigma_x$ and $\sigma_y$ in the Bloch Hamiltonian $H_0(\boldsymbol k)$, which means that the condition for the existence of a disorder-insensitive Bloch mode is complex. Thus, their codimension is two and they will only appear at certain isolated momenta. This is confirmed by the numerical studies we conducted on disordered Honeycomb models, where Bloch modes survived at isolated momenta under the presence of lattice-selective disorder but no EPs where observed. However, lattice-selective disorder in graphene has been linked to other interesting phenomena by a previous study \cite{Graphene_1, Graphene_2}.

\subsection{Ultracold atoms in optical lattices with spin-selective disorder}
In the context of atomic many-body systems, a promising platform for experimentally implementing our model system is provided by ultracold atoms in (spin-dependent) optical lattices \cite{Review_UCA}. In the quest for probing many-body localization, the experimental realization and control of disorder potentials in such synthetic material systems in both 1D and 2D has been impressively demonstrated \cite{UCA_MB_Localization_1, UCA_MB_Localization_2}. Moreover, quantum gas microscopy methods with spin-selective single site resolution have enabled the observation of anti-ferromagnetic phases with cold atoms in optical lattices \cite{UCA_Microscope_1, UCA_Microscope_2,
UCA_AFM}. Combining these recent developments, our proposed 2D Dirac models with spin-selective disorder potential are well within the state of the art toolbox of atomic quantum simulators. Remarkably, the enormous experimental control over such systems does not only allow for the observation of spectral properties: Despite the electrically neutral character of the atoms, also two terminal transport setups have become amenable in the laboratory \cite{UCA_Transport_1, UCA_Transport_2}.

\subsection{Surface states of a topological insulator}
Another experimentally well studied platform for 2D Dirac physics is based on the surface states of a time-reversal symmetric topological insulator (TI) exposed to magnetic perturbations that affect the electron spin representing the internal degree of freedom in this setting. There, an exceptional NH phase has previously been proposed to be induced by the tunnel coupling to a ferromagnetic thermal reservoir \cite{Bergholtz2019}. These ideas may be adapted to our present context of NH physics induced by disorder scattering in a straight-forward way. Instead of a magnetic lead, the relevant perturbation is then given by magnetic impurity ions adsorbed to the surface such that the electro-static potential generates the $s_0 \sigma_0$ disorder and the magnetic moment the $\boldsymbol s  \cdot \boldsymbol \sigma$ term (cf.~Eq.~(\ref{eqn:disorderingredients})). However, it is fair to say that several issues remain with the disordered version of this setting. To satisfy ingredient $(ii)$, fine-tuning the amplitude of $s_0$ and $\boldsymbol s $ is required. Furthermore, the above ingredient $(iii)$ requires that the on-site potential $s_0 \sigma_0$ changes sign with the direction of the magnetic moment $\boldsymbol s  \cdot \boldsymbol \sigma$. This amounts to having positive and negative magnetic ions with oppositely oriented magnetic moments.

\subsection{Comparison to materials with tilted Dirac cones}
Previous works have demonstrated that EPs can emerge from trivial disorder (i.e. affecting all degrees of freedom similarly) as well in two-dimensional Dirac materials with tilted cones \cite{EPDisorder1, EPDisorder2, EPDisorder3}. While such systems are probably easier to fabricate than the ones listed so far, the resulting Fermi arcs are unfortunately buried under a  finite imaginary part that increases with the size of the Fermi arcs, since the non-decaying quasiparticles can only be obtained through orbital-restricted disorder. This is likely to obscure transport signatures and renders this class of systems less interesting for our present purposes.

\section{Conclusion}
In this work, we have demonstrated the occurence of disorder-induced EPs by means of exact numerical calculations and explored their impact on the transport properties of a finitely-sized sample. For this, we study a minimal model of a Dirac semimetal, which can enter a conventional phase (without EPs) or an exceptional phase through the addition of random disorder on both or predominantly on one of the sublattices. The exceptional phase exhibits quasiparticle excitations with prolonged lifetime that we tie to orbital restricted disorder through a rigorous theorem. Besides studying the spectral properties, we compute the two-terminal transmission, and find that a finite sample in the exceptional phase shows greatly enhanced transport properties at zero energy when compared to both the clean system and the conventional disordered phase. We demonstrate that this effect is carried by the aforementioned long-lived quasiparticles, and accompanied by an anomalously slow decay of the localized eigenfunctions. Finally, we outline possible platforms for the experimental realization of the proposed model systems, where the main challenge is represented by engineering orbital-selective disorder.
 
\acknowledgments
{\it Acknowledgments.---}
We would like to thank T. Micallo and M. Vojta for discussions. We acknowledge financial support from the German Research Foundation (DFG) through the Collaborative Research Centre SFB 1143, the Cluster of Excellence ct.qmat, and the DFG Project 419241108. Our numerical calculations were performed on resources at the TU Dresden Center for Information Services and High Performance Computing (ZIH).

\onecolumngrid

\appendix

\section{Perturbation Theory}\label{perturbation_theory}
Here, we will discuss how the disorder-averaging restores translational invariance, leading to a self-energy that is local in $\bm k$, and derive a perturbative expression for the self-energy correction on the basis of chapter 12 of \cite{S_MBQT}. In the following, we use imaginary frequency $i \omega$ to shorten the notation and replace it with $i \omega \to \omega + i \eta$ at the end.

For a system consisting of a solvable Hamiltonian $H_0$ plus some perturbation $V$, such that the full Hamiltonian is $H = H_0 + V$ and the free retarded Green's function (GF) $G^0(i \omega) = [\mathbbm{1}i \omega - H_0]^{-1}$ in frequency space is known, one can self-insert the Dyson equation to obtain a perturbation series for the full retarded GF
$G(i \omega) = [\mathbbm{1}i \omega - H]^{-1} = [\mathbbm{1}i \omega - H_0 - V]^{-1}$. The result is

\begin{align}
G(i \omega) &= G^0(i \omega)+ G^0(i \omega)V G^0(i \omega)
+  G^0(i \omega)V  G^0(i \omega)V  G^0(i \omega) + ... \label{PTB_series}
\end{align} 

If the unperturbed system $H_0$ is a tight-binding model with translational invariance and  $n$ internal degrees of freedom on each site, the free GF is block-diagonal in the Bloch basis and reads 
$G^0_{\bm k, \bm k'}(i \omega) =  \delta_{\bm k, \bm k'} G^0(\bm k, i \omega)$ where $G^0(\bm k, i \omega)=  [\mathbbm{1}i \omega- H_0( \bm k)]^{-1}$ and $H_0(\bm k)$ is the $n$x$n$-Bloch-Hamiltonian.

We will consider the case of random disorder represented by similar impurities at each site, but with uncorrelated random amplitudes $a_{\bf j}$ (with $\boldsymbol j \in  \{1, 2, ...,N_\mathrm{x}\} \times \{1, 2, ...,N_\mathrm{y}\}$) distributed according to a distribution function $f(a)$. The disorder term $V$ may include on-site terms and hoppings between sites connected by a vector $\boldsymbol \delta = (\delta_\mathrm{x} , \delta_\mathrm{y} )$. The generic form  in real space representation is then

\begin{align}
V =&  \sum_{\boldsymbol j }  a_{\boldsymbol j}  \left [\frac{1}{2} \Psi_{\boldsymbol j}^\dagger V_\mathrm{OS} \Psi_{\boldsymbol j} 
+ \Psi_{\boldsymbol j+\boldsymbol \delta_1}^\dagger V_\mathrm{\boldsymbol \delta_1} \Psi_{j} + \Psi_{\boldsymbol j+\boldsymbol \delta_2}^\dagger V_\mathrm{\boldsymbol \delta_2} \Psi_{j} + ... + \mathrm{h.c.} \right] \nonumber \\
=& \sum_{\boldsymbol j }  \sum_{\boldsymbol k, \boldsymbol k'} a_{\boldsymbol j} \frac{e^{-i \boldsymbol j \cdot (\boldsymbol k- \boldsymbol k')}}{N_\mathrm{x}  N_\mathrm{y}} c_{\boldsymbol k}^\dagger
\underbrace{\left[V_\mathrm{OS} + ( V_\mathrm{\boldsymbol \delta_1} e^{-i \boldsymbol \delta_1 \cdot \boldsymbol k} 
+ V_\mathrm{\boldsymbol \delta_1}^\dagger e^{i \boldsymbol \delta_1 \cdot \boldsymbol k'}) + ( V_\mathrm{\boldsymbol \delta_2} e^{-i \boldsymbol \delta_2 \cdot \boldsymbol k} 
+ V_\mathrm{\boldsymbol \delta_2}^\dagger e^{i \boldsymbol \delta_2 \cdot \boldsymbol k'})
+ ... \right ]}_{V_{\boldsymbol k, \boldsymbol k'}} c_{\boldsymbol k'}. \label{impurity}
\end{align}

The $\Psi_{\bm j}^\dagger$ are $n$-spinors of creators for on-site states and
the $c_{\bm k}^\dagger = \frac{1}{\sqrt{N_\mathrm{x} N_\mathrm{y}}}\sum_{\bm j} e^{i \bm j \cdot \bm k} \Psi_{\bm j}^\dagger$ are $n$-spinors of creators for the Bloch-states. The series from 
Eq.~(\ref{PTB_series}) leads to an expression for the blocks of the full GF

\begin{align}
G_{\bm k, \bm k'}(i \omega) =  &\delta_{\bm k, \bm k'} G^0(\bm k, i \omega) + \sum_{\bm j_1}  G^0(\bm k, i \omega)
a_{\bm j_1}\frac{e^{-i\bm j_1 \cdot (\bm k - \bm k')}}{N_\mathrm{x} N_\mathrm{y}} V_{\bm k,\bm k'} G^0(\bm k', i \omega) \nonumber \\
& + \sum_{\bm q}\sum_{\bm j_1,\bm j_2 } G^0(\bm k, i \omega)
a_{\bm j_1}\frac{e^{-i \bm j_1 \cdot (\bm k - \bm q)}}{N_\mathrm{x} N_\mathrm{y}} V_{\bm k,\bm q} G^0(\bm q, i \omega)
a_{\bm j_2}\frac{e^{-i \bm j_2 \cdot (\bm q - \bm k')}}{N_\mathrm{x} N_\mathrm{y}} V_{\bm q,\bm k'} G^0(\bm k', i \omega) +... \nonumber
\end{align}
To obtain the effective Hamiltonian, this expression is averaged over the impurity amplitudes $\{a_{ \bm j } \}$ by calculating
\begin{align}
G_{\bm k, \bm k'}^{av}(i \omega) = \langle G_{\bm k,\bm k'}(i \omega) \rangle_f = \int \mathrm{d}a_{1,1} f(a_{1,1}) 
\int \mathrm{d}a_{1,2} f(a_{1,2}) ... \int \mathrm{d}a_{N_\mathrm{x} ,N_\mathrm{y}} f(a_{N_\mathrm{x} ,N_\mathrm{y}})
G_{\bm k,\bm k'}(i \omega). \label{G_av} 
\end{align} 
In Eq.~(\ref{G_av}) we encounter terms of the form 
$\langle \sum_{\bm j_1, ...\bm j_m } a_{\bm j_1}a_{\bm j_2}...a_{\bm j_{m}}e^{\sum^m_{l=1}\bm  q_l \cdot \bm j_l}\rangle_f$. We can group the sums into those where all scattering 
vectors $\bm q \in \mathbf{Q} = \{\bm q_1, \bm q_2, ...,\bm q_m\}$ are connected to one, two, three and so on impurities. The notation $|\mathbf{Q}_r|$ simply indicates the number of elements in the subset $\mathbf{Q}_r \subset \mathbf{Q} $. 
\begin{align}
\langle \sum_{\bm j_1, ...\bm j_m } a_{\bm j_1}a_{\bm j_2}...a_{\bm j_{m}}e^{\sum^m_{l=1}\bm  q_l \cdot \bm j_l}\rangle_f= &
\langle \sum_{\bm h_1=1} (a_{\bm h_1})^m e^{\sum_{\bm q \in \mathbf{Q}} \bm q \cdot\bm h_1}\rangle_f \nonumber \\
&+ \langle \sum_{\cup^2_{r = 1} \mathbf{Q}_r = \mathbf{Q}} \sum_{\bm h_1} 
\sum_{\substack{\bm h_2 \\ \bm h_2 \neq \bm h_1}}
(a_{\bm h_1})^{|\mathbf{Q}_1|}(a_{\bm h_2})^{|\mathbf{Q}_2|} e^{\sum_{\bm q_1 \in \mathbf{Q_1}} \bm q_1 \cdot  \bm h_1} 
e^{\sum_{\bm q_2 \in \mathbf{Q_2}} \bm q_2 \cdot \bm h_2}\rangle_f \nonumber \\
&+ \langle \sum_{\cup^3_{r = 1} \mathbf{Q}_r = \mathbf{Q}} \sum_{\bm h_1} 
\sum_{\substack{\bm h_2\\ \bm h_2 \neq \bm h_1}} \sum_{\substack{\bm h_3 \\ \bm h_3 \neq \bm h_1,\bm h_2}} (a_{\bm h_1})^{|\mathbf{Q}_1|}(a_{\bm h_2})^{|\mathbf{Q}_2|}(a_{\bm h_3})^{|\mathbf{Q}_3|}  \nonumber \\
& \times e^{\sum_{\bm q_1 \in \mathbf{Q_1}} \bm q_1 \cdot h_1}
e^{\sum_{\bm q_2 \in \mathbf{Q_2}} \bm q_2 \cdot \bm h_2} e^{\sum_{\bm q_3 \in \mathbf{Q_3}} \bm q_3 \cdot \bm h_3}\rangle_f \nonumber\\
&+ ... \nonumber
\end{align}
Now we need to introduce a small error of the order $\frac{1}{N_\mathrm{x} N_\mathrm{y}}$ by letting the $\bm h$ 
sums run unrestricted, e.g. 
$\sum_{\substack{\bm h_2 \\ \bm h_2 \neq \bm h_1}} \to \sum_{\bm h_2 }$. The average doesn't act on the exponentials and the distribution of the amplitudes is uncorrelated, so we can pull the average of the amplitudes out of the sums. Performing the sums gives a delta function for all momenta connected to the same impurity and we arrive at

\begin{align}
\langle \sum_{\bm j_1, ...\bm j_m } a_{\bm j_1}a_{\bm j_2}...a_{j_{m}}e^{\sum^m_{l=1} q_l j_l}\rangle_f = &
N_\mathrm{x} N_\mathrm{y} \langle a^m \rangle_f \delta_{0,\sum_{\bm q \in \mathbf{Q}}\bm q} \nonumber \\
&+ (N_\mathrm{x} N_\mathrm{y})^2\sum_{\cup^2_{r = 1} \mathbf{Q}_r = \mathbf{Q} }\delta_{0,\sum_{\bm q_1 \in \mathbf{Q_1}}\bm q_1}
\langle a^{|\mathbf{Q}_1|} \rangle_f \langle a^{|\mathbf{Q}_2|} \rangle_f
 \delta_{0,\sum_{\bm q_2 \in \mathbf{Q_2}}\bm q_2} \nonumber \\
&+ (N_\mathrm{x} N_\mathrm{y})^3\sum_{\cup^3_{r = 1} \mathbf{Q}_r = \mathbf{Q}} \langle a^{|\mathbf{Q}_1|} \rangle_f
\langle a^{|\mathbf{Q}_2|} \rangle_f \langle a^{|\mathbf{Q}_3|} \rangle_f
\delta_{0,\sum_{\bm q_1 \in \mathbf{Q_1}}\bm q_1}\delta_{0,\sum_{\bm q_2 \in \mathbf{Q_2}}\bm q_2} \delta_{0,\sum_{\bm q_3 \in \mathbf{Q_3}}\bm q_3} \nonumber \\
&+... \nonumber
\end{align}

Bearing this result in mind, we can express Eq.~(\ref{G_av}) in terms of Feynman diagrams. It is given by the sum over all topologically different diagrams of the form

\begin{align}
G_{\bm k,\bm k}^{av}(i \omega)=
\vcenter{\hbox{
\begin{tikzpicture}
  \begin{feynman}
     \vertex (e1) at (-1, 0);
     \vertex (e2) at (0, 0);
     \vertex (i1) at (0, 4)[]{};
     \vertex [above=1em of i1, opacity = 0] {\(\langle a \rangle_f\)};
     \vertex [below=0.5em of e2, opacity = 0] {\(V_{\bm k,\bm k}\)};
    \diagram*{
    (e2) -- [fermion, edge label=\(\bm k\)] (e1),
    (i1) -- [charged scalar, opacity = 0] (e2),
    };
  \end{feynman}
\end{tikzpicture}}}
+
\vcenter{\hbox{
\begin{tikzpicture}
  \begin{feynman}
     \vertex (e1) at (-1, 0);
     \vertex (e2) at (1, 0);
     \vertex (a) at (0, 0);
     \vertex (i1) at (0, 4)[crossed dot]{};
     \vertex [above=1em of i1] {\(\langle a \rangle_f\)};
     \vertex [below=0.5em of a] {\(V_{\bm k,\bm k}\)};
    \diagram*{
    (a) -- [charged scalar, edge label'=\(0\)] (i1),
    (e2) -- [fermion, edge label=\(\bm k\)] (a),
    (a) -- [fermion, edge label=\(\bm k\)] (e1),
    };
  \end{feynman}
\end{tikzpicture}}}
+
\vcenter{\hbox{
\begin{tikzpicture}
  \begin{feynman}
     \vertex (e1) at (-1, 0);
     \vertex (e2) at (2, 0);
     \vertex (a) at (0, 0);
     \vertex (i1) at (0.5, 4)[crossed dot]{};
     \vertex (b) at (1, 0);
     \vertex [above=1em of i1] {\(\langle a^2 \rangle_f\)};
     \vertex [below=0.5em of a] {\(V_{\bm k,\bm q}\)};
     \vertex [below=0.5em of b] {\(V_{\bm q,\bm k}\)};
    \diagram*{
      (b) -- [fermion, edge label=\(q\)] (a) ,
      (a) -- [charged scalar, edge label=\(\bm q- \bm k\)] (i1),
      (b) -- [charged scalar, edge label'=\(\bm k-\bm q\)] (i1),
      (e2) -- [fermion, edge label=\(\bm k\)] (b),
      (a) -- [fermion, edge label=\(\bm k\)] (e1),
    };
  \end{feynman}
\end{tikzpicture}}}
+
\vcenter{\hbox{
\begin{tikzpicture}
  \begin{feynman}
     \vertex (e1) at (-1, 0);
     \vertex (e2) at (2, 0);
     \vertex (a) at (0, 0);
     \vertex (b) at (1, 0);
     \vertex (i1) at (0, 4)[crossed dot]{};
     \vertex (i2) at (1, 4)[crossed dot]{};
     \vertex [above=1em of i1] {\(\langle a \rangle_f\)};
     \vertex [below=0.5em of a] {\(V_{\bm k,\bm k}\)};
     \vertex [above=1em of i2] {\(\langle a \rangle_f\)};
     \vertex [below=0.5em of b] {\(V_{\bm k,\bm k}\)};
    \diagram*{
    (a) -- [charged scalar, edge label'=\(0\)] (i1),
    (b) -- [charged scalar, edge label'=\(0\)] (i2),
    (b) -- [fermion, edge label=\(\bm k\)] (a),
    (e2) -- [fermion, edge label=\(\bm k\)] (b),
    (a) -- [fermion, edge label=\(\bm k\)] (e1),
    };
  \end{feynman}
\end{tikzpicture}}}
+... , \label{G_av_diagram}
\end{align}
which obey simple Feynman rules. The solid-lined propagators with momentum $\bm k$ denote a matrix-valued free GF $G^0(\bm k, i \omega)$. The dashed propagators denote an also matrix-valued factor $V_{\bm q_\mathrm{L},\bm q_\mathrm{R}}$, where $\bm q_\mathrm{L}$ is the momentum leaving the vertex of the dashed and the two solid propagators to the left and $\bm q_\mathrm{R}$ the momentum joining it from the right. A vertex of $m$ dashed propagators obtains the $m$-th moment of the distribution $\langle a^m \rangle_f$ as a prefactor. The dashed propagators formally carry the momentum $\bm q_\mathrm{R} - \bm q_\mathrm{L}$ and all momenta joining a vertex of multiple dashed propagators add up to zero.  A sum $\frac{1}{N_\mathrm{x} N_\mathrm{y}}\sum_{\bm q}$ over all momenta inside a closed loop is implied.

Now the series can be rearranged by defining the self-energy $\Sigma(\bm k, i\omega)$ as the sum of all irreducible diagrams that cannot be separated by cutting a single propagator

\begin{align}
G_{k,k}^{av}(i \omega)&=
\vcenter{\hbox{
\begin{tikzpicture}
  \begin{feynman}
     \vertex (e1) at (-0.5, -0.5);
     \vertex (e2) at (0.5, -0.5);
    \diagram*{
    (e2) -- [fermion, edge label=\(\bm k\)] (e1),
    (e2) -- [fermion, half left, opacity = 0] (e1),
    (e1) -- [fermion, half left, opacity = 0] (e2),
    };
  \end{feynman}
\end{tikzpicture}}}
+
\vcenter{\hbox{
\begin{tikzpicture}
  \begin{feynman}
     \vertex (e1) at (-1, 0);
     \vertex (e2) at (2, 0);
     \vertex (a) at (0, 0);
     \vertex (b) at (1, 0);
     \vertex [right=0.7em of a] {\(\Sigma\)};
    \diagram*{
    (b) -- [fermion,half left] (a),
    (a) -- [fermion, half left] (b),
    (e2) -- [fermion, edge label=\(\bm k\)] (b),
    (a) -- [fermion, edge label=\(\bm k\)] (e1),
    };
  \end{feynman}
\end{tikzpicture}}}
+
\vcenter{\hbox{
\begin{tikzpicture}
  \begin{feynman}
     \vertex (e1) at (-1, 0);
     \vertex (e2) at (4, 0);
     \vertex (a) at (0, 0);
     \vertex (b) at (1, 0);
     \vertex (c) at (2, 0);
     \vertex (d) at (3, 0);
     \vertex [right=0.7em of a] {\(\Sigma\)};
     \vertex [right=0.7em of c] {\(\Sigma\)};
    \diagram*{
    (b) -- [fermion,half left] (a),
    (a) -- [fermion, half left] (b),
    (d) -- [fermion,half left] (c),
    (c) -- [fermion, half left] (d),
    (e2) -- [fermion, edge label=\(\bm k\)] (d),
    (c) -- [fermion, edge label=\(\bm k\)] (b),
    (a) -- [fermion, edge label=\(\bm k\)] (e1),
    };
  \end{feynman}
\end{tikzpicture}}}
+... \nonumber \\
&=  G^0(\bm k, i \omega) + G^0(\bm k, i \omega) \Sigma(\bm k, i\omega)\left[G^0(\bm k, i \omega) + G^0(\bm k, i \omega) \Sigma(\bm k, i\omega) G^0(\bm k, i \omega) + ...\right] \nonumber\\
&=  G^0(\bm k, i \omega) + G^0(\bm k, i \omega) \Sigma(\bm k, i\omega)\left[G_{\bm k,\bm k}^{av}(i \omega)\right]. \nonumber
\end{align}
Because the free Green's function is given by $G^0(k, i \omega)=  [\mathbbm{1}i \omega- H_0(\bm k)]^{-1}$, we finally obtain

\begin{align}
G_{\bm k,\bm k}^{av}(i \omega) = \left[\mathbbm{1} - \left(H_0(\bm k) + \Sigma(\bm k, i\omega)\right) \right]^{-1}. \nonumber 
\end{align}
As was said, the self-energy consists of all irreducible diagrams and assumes the form

\begin{align}
\Sigma(\boldsymbol k, i\omega)=
\vcenter{\hbox{
\begin{tikzpicture}
  \begin{feynman}
     \vertex (a) at (0, 0);
     \vertex (i1) at (0, 4)[crossed dot]{};
     \vertex [above=1em of i1] {\(\langle a \rangle_f\)};
     \vertex [below=0.5em of a] {\(V_{\boldsymbol k,\boldsymbol k}\)};
    \diagram*{
    (a) -- [charged scalar, edge label'=\(0\)] (i1),
    };
  \end{feynman}
\end{tikzpicture}}}
+
\vcenter{\hbox{
\begin{tikzpicture}
  \begin{feynman}
     \vertex (a) at (0, 0);
     \vertex (i1) at (1.5, 4)[crossed dot]{};
     \vertex (b) at (3, 0);
     \vertex [above=1em of i1] {\(\langle a^2 \rangle_f\)};
     \vertex [below=0.5em of a] {\(V_{\boldsymbol k,\boldsymbol q}\)};
     \vertex [below=0.5em of b] {\(V_{\boldsymbol q,\boldsymbol k}\)};
    \diagram*{
      (b) -- [fermion, edge label=\(\boldsymbol q\)] (a) ,
      (a) -- [charged scalar, edge label=\(\boldsymbol q-\boldsymbol k\)] (i1),
      (b) -- [charged scalar, edge label'=\(\boldsymbol k- \boldsymbol q\)] (i1),
    };
  \end{feynman}
\end{tikzpicture}}}
+
\vcenter{\hbox{
\begin{tikzpicture}
  \begin{feynman}
     \vertex (a) at (0, 0);
     \vertex (i1) at (1.5, 4)[crossed dot]{};
     \vertex (i2) at (1.5, 2)[crossed dot]{};
     \vertex (b) at (1.5, 0);
     \vertex (c) at (3, 0);
     \vertex [above=1em of i1] {\(\langle a^2 \rangle_f\)};
     \vertex [above=1em of i2] {\(\langle a \rangle_f\)};
     \vertex [below=0.5em of a] {\(V_{\boldsymbol k,\boldsymbol q}\)};
     \vertex [below=0.5em of b] {\(V_{\boldsymbol q,\boldsymbol q}\)};
     \vertex [below=0.5em of c] {\(V_{\boldsymbol q,\boldsymbol k}\)};
    \diagram*{
      (b) -- [fermion, edge label=\(\boldsymbol q\)] (a) ,
      (a) -- [charged scalar, edge label=\(\boldsymbol q- \boldsymbol k\)] (i1),
      (b) -- [charged scalar, edge label'=\(0\)] (i2),
      (c) -- [charged scalar, edge label'=\(\boldsymbol k - \boldsymbol q\)] (i1),
      (c) -- [fermion, edge label=\(\boldsymbol q\)] (b) ,
    };
  \end{feynman}
\end{tikzpicture}}}
+...  \label{SE_Full} 
\end{align}

For on-site impurities, i.e. impurities without hopping terms, Eq.~(\ref{impurity}) shows that $V_{\boldsymbol k, \boldsymbol k'}$ is a constant and thus the self-energy $\Sigma(\boldsymbol k, \omega + i \eta)$ does not depend on $\bm k$ as well, since we integrate over all internal momenta.

\section{Travelling Distance of Modes Through the Sample} \label{travelling_distance}

\begin{wrapfigure}[3]{r}{5cm}  
        \vspace{-10pt}
        \centering
        \includegraphics[width= 5cm ]{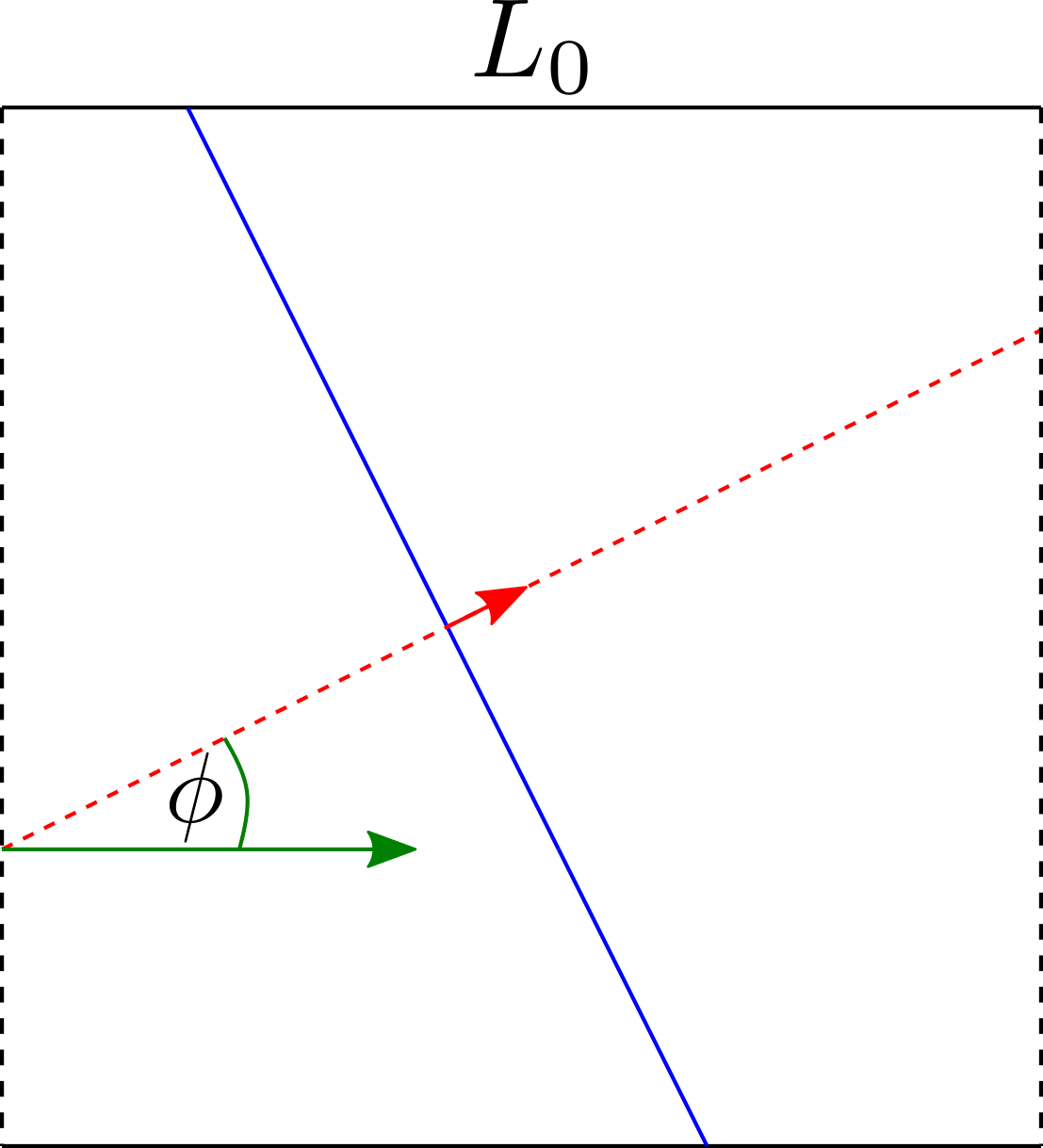}
        \caption{A wave front (in blue) propagates through the sample at an angle $\phi$.}  
        \label{travel_distance}      
\end{wrapfigure}

Fig.~\ref{travel_distance} shows how a wave front travels through a sample of length $L_0$ at an angle $\phi$. The red line represents the distance that each point of the wave front has to travel. 
From basic trigonometry, one can estimate the length of the red line to be

\[\centering
L = L_0  \sqrt{1 + \tan(\phi)^2}.
\]

\vspace{120pt}

\section{Disorder-insensitive Bloch modes} \label{APP:Bloch_States}

\subsection{Prove of the theorem}
Consider a generic tight binding Hamiltonian $H_0$ with $n$ internal degrees of freedom and translational invariance perturbed by a random disorder term $V$ that only affects a subset $\{ \alpha \}$ of the orbitals in the sense of random on-site potentials as well as random hopping terms connecting $\{ \alpha \}$ orbitals from different unit cells. We will denote the complement of $\{ \alpha \}$ by $\{ \beta \}$, the number of disorder affected orbitals by $n_\alpha$ and the number of the remaining orbitals by $n_\beta = n - n_\alpha$.

Assuming periodic boundary conditions or the thermodynamic limit, we may diagonalize $H_0$ by switching to the Bloch basis with the transformation

\begin{equation}
\Psi_{\bm j, \alpha} = \frac{1}{\sqrt{N}} \sum_{\bm k} e^{i \bm k (\bm r_{\bm j} + \bm  r_\alpha)} c_{\bm k, \alpha}, \nonumber \\
\Psi_{\bm j, \beta} = \frac{1}{\sqrt{N}} \sum_{\bm k} e^{i \bm k (\bm r_{\bm j} + \bm  r_\beta)} c_{\bm k, \beta} \nonumber
\end{equation} 
where $\Psi_{\bm j, \alpha} $ ($\Psi_{\bm j, \beta} $) are the field operators annihilating an electron in orbital $\alpha$ ($\beta$) in unit cell $\bm j$, $\bm r_{\bm j}$ is the position of that unit cell, $\bm r_\alpha$ ($\bm r_\beta$) are the positions of the orbital inside that unit cell and $N$ is the number of sites. In the Bloch basis, $H_0$ can then be written as 
\begin{equation}
H_0 = \sum_{\bm k} c^\dagger_{\bm k, \bm \alpha} \mathcal{H}_{0,\alpha}(\bm k) c_{\bm k, \bm \alpha} +  \sum_{\bm k} c^\dagger_{\bm k, \bm \beta} \mathcal{H}_{0,\beta}(\bm k) c_{\bm k, \bm \beta}  + \sum_{\bm k} c^\dagger_{\bm k, \bm \alpha} \mathcal{H}_{0,\alpha,\beta}(\bm k) c_{\bm k, \bm \beta} 
+ c^\dagger_{\bm k, \bm \beta} \left( \mathcal{H}_{0,\alpha,\beta}(\bm k) \right)^\dagger c_{\bm k, \bm \alpha}, \label{H_0_Bloch_Basis}
\end{equation} 
where $c^\dagger_{\bm k, \bm \alpha}$, $c^\dagger_{\bm k, \bm \beta}$ denote the vectors of all Bloch wave state createors belonging to the set $\{ \alpha \}$, $\{ \beta \}$, $\mathcal{H}_{0,\alpha}(\bm k)$ is an $n_\alpha \times n_\alpha$ matrix, $ \mathcal{H}_{0,\beta}(\bm k)$ is an $n_\beta \times n_\beta$ matrix 
and $\mathcal{H}_{0,\alpha,\beta}(\bm k)$ is an $n_\alpha \times n_\beta$ matrix. 

The random disorder term $V$ takes the form 

\begin{equation}
V = \sum_{\bm k, \bm k'} c^\dagger_{\bm k, \bm \alpha} V_{\bm k, \bm k'} c_{\bm k, \bm \alpha} + h.c.  \label{V_Bloch_Basis}
\end{equation} 
with a matrix $V_{\bm k, \bm k'}$ of size $n_\alpha \times n_\alpha$.

In order to find a Bloch state that lives entirely on the subset of orbitals  $\{ \beta \}$, we may perform a unitary transformation $c_{\bm k, \bm \beta}  \to c'_{\bm k, \bm \beta} $ that diagonalizes  $\mathcal{H}_{0,\beta}(\bm k)$ such that Eq.~(\ref{H_0_Bloch_Basis}) becomes

\begin{equation}
H_0 = \sum_{\bm k} c^\dagger_{\bm k, \bm \alpha} \mathcal{H}_{0,\alpha}(\bm k) c_{\bm k, \bm \alpha} +  \sum_{\bm k} \sum_{\beta} \varepsilon_{\beta'}(\bm k)c'^\dagger_{\bm k, \beta} c'_{\bm k,  \beta}  + \sum_{\bm k} c^\dagger_{\bm k, \bm \alpha} \mathcal{H'}_{0,\alpha,\beta}(\bm k) c'_{\bm k, \bm \beta} 
+ c'^\dagger_{\bm k, \bm \beta} \left( \mathcal{H'}_{0,\alpha,\beta}(\bm k) \right)^\dagger c_{\bm k, \bm \alpha}. \label{H_0_Bloch_Basis_Transformed}
\end{equation} 
In order for $|\bm k_0, \beta' \rangle = c'^\dagger_{\bm k_0, \beta} |0\rangle$ to be an eigenstate of $H_0$ for some $\bm k_0$, the corresponding column of the coupling matrix $ \mathcal{H'}_{0,\alpha,\beta}(\bm k_0)$ has to vanish. These are $n_\alpha$ complex conditions for the existence of an eigenstate to $H_0$ with energy $\varepsilon_{\beta'}(\bm k_0)$ that solely lives on the sublattice $\{ \beta \}$. 

From Eq.~(\ref{V_Bloch_Basis}) and Eq.~(\ref{H_0_Bloch_Basis_Transformed}), it is evident that $|\bm k_0, \beta' \rangle $ is also an eigenstate to the total system $H = H_0 + V$

\begin{equation}
 (H_0 + V)|\bm k_0, \beta' \rangle = (H_0 + V)c'^\dagger_{\bm k_0, \beta} |0\rangle = \varepsilon_{\beta'}(\bm k_0)|\bm k_0, \beta' \rangle .\nonumber
\end{equation} 

To conclude, we find the codimension of disorder-insensitive Bloch modes for a generic system with $n_\alpha$ disorder-affected orbitals to be $2 n_\alpha$. However, certain symmetries may turn the $n_\alpha$ complex conditions to real ones or require that some of the entries of $ \mathcal{H'}_{0,\alpha,\beta}(\bm k)$ vanish, thereby reducing the codimension of these Bloch states.

\subsection{Application to our model}
The unperturbed Hamiltonian $H_0$ from Eq.~(\ref{H_0}) obeys the chiral symmetry

\begin{align}
\sigma_y H_0 \sigma_y = - H_0 \label{symmetry}
\end{align}
In general, this will allow for a Bloch Hamiltonian of the form
\begin{align}
H_0 = \sum_{\bm k} (c^\dagger_{\bm k, A}, c^\dagger_{\bm k, B}) (h_{\sigma_x}(\bm k) + h_{\sigma_z}(\bm k) )  (c_{\bm k, A}, c_{\bm k, B})^\mathrm{T} = 
\sum_{\bm k} (c^\dagger_{\bm k, A}, c^\dagger_{\bm k, B}) 
\begin{pmatrix}
h_{\sigma_z}(\bm k) && h_{\sigma_x}(\bm k)\\
h_{\sigma_x}(\bm k) && - h_{\sigma_z}(\bm k)
\end{pmatrix}
\begin{pmatrix}
c_{\bm k, A}\\
 c_{\bm k, B}
\end{pmatrix}. \label{H_0_Bloch}
\end{align}
The disorder term $V$ from Eq.~(\ref{disorder}) can be brought to a form that only affects sublattice $A$ by performing a unitary transformation that amounts to a rotation around the $y$ axis in the space of Pauli matrices and leaves the symmetry from Eq.~(\ref{symmetry}) invariant. The decomposition of $H_0$ from Eq.~(\ref{H_0_Bloch}) in the sense of Eq.~(\ref{H_0_Bloch_Basis_Transformed}) yields

\begin{align}
\mathcal{H}_{0,\alpha}(\bm k) = h_{\sigma_z}(\bm k), \nonumber \\
\varepsilon_\beta(\bm k_0) = - h_{\sigma_z}(\bm k), \nonumber \\
\mathcal{H}_{0,\alpha, \beta}(\bm k) = h_{\sigma_x}(\bm k). \nonumber
\end{align}
Note that $\mathcal{H}_{0,\alpha, \beta}(\bm k)$ must be real and that this is a consequence of the chiral symmetry from Eq.~(\ref{symmetry}). The condition for the existence of a disorder-insensitive Bloch mode with some momentum $\bm k_0$ is now
\begin{align}
h_{\sigma_x}(\bm k_0) = 0. \nonumber
\end{align}
and the associated eigenenergy is
\begin{align}
\varepsilon_\beta(\bm k_0) = -h_{\sigma_z}(\bm k_0). \nonumber
\end{align}

\end{document}